\documentclass[12pt,a4paper]{article}
\usepackage[utf8]{inputenc}
\usepackage[T1]{fontenc}
\usepackage{amsmath, amssymb, amsthm}
\usepackage{graphicx}
\usepackage{hyperref}
\usepackage{geometry}
\usepackage{booktabs}
\usepackage{mathtools}
\usepackage{dsfont}
\usepackage{xcolor}
\usepackage{enumitem}
\usepackage{float}
\usepackage{setspace}
\usepackage{tikz}
\usepackage{pgfplots}
\pgfplotsset{compat=1.15}
\usetikzlibrary{arrows.meta, positioning, patterns, calc, matrix}
\newtheorem{proposition}{Proposition}[section]
\newtheorem{theorem}[proposition]{Theorem}

\newtheorem{remark}[proposition]{Remark}

\newtheorem{axiom}{Axiom}

\newcommand{\Ent}{\mathcal{E}}
\newcommand{\W}{\mathcal{W}}
\newcommand{\F}{\mathcal{F}}
\newcommand{\Om}{\Omega}
\newcommand{\dd}{\,\mathrm{d}}

\newcommand{\humrich}[3]{%
  \fill[#3] (#1,#2+0.35) circle (0.09);%
  \draw[#3, line width=1.8pt] (#1,#2+0.26) -- (#1,#2-0.15);%
  \draw[#3, line width=1.8pt] (#1-0.12,#2+0.10) -- (#1+0.12,#2+0.10);%
  \draw[#3, line width=1.8pt] (#1,#2-0.15) -- (#1-0.08,#2-0.4);%
  \draw[#3, line width=1.8pt] (#1,#2-0.15) -- (#1+0.08,#2-0.4);%
  \fill[#3!50] (#1+0.14,#2-0.08) rectangle (#1+0.24,#2+0.02);%
}
\newcommand{\humpoor}[3]{%
  \fill[#3] (#1,#2+0.22) circle (0.075);%
  \draw[#3, line width=1.4pt] (#1,#2+0.14) -- (#1,#2-0.08);%
  \draw[#3, line width=1.4pt] (#1-0.09,#2+0.03) -- (#1+0.09,#2+0.03);%
  \draw[#3, line width=1.4pt] (#1,#2-0.08) -- (#1-0.06,#2-0.25);%
  \draw[#3, line width=1.4pt] (#1,#2-0.08) -- (#1+0.06,#2-0.25);%
}
\newcommand{\hummid}[3]{%
  \fill[#3] (#1,#2+0.28) circle (0.08);%
  \draw[#3, line width=1.6pt] (#1,#2+0.20) -- (#1,#2-0.10);%
  \draw[#3, line width=1.6pt] (#1-0.10,#2+0.06) -- (#1+0.10,#2+0.06);%
  \draw[#3, line width=1.6pt] (#1,#2-0.10) -- (#1-0.07,#2-0.30);%
  \draw[#3, line width=1.6pt] (#1,#2-0.10) -- (#1+0.07,#2-0.30);%
}
\newcommand{\humrichNB}[3]{\fill[#3] (#1,#2+0.35) circle (0.09);\draw[#3,line width=1.8pt](#1,#2+0.26)--(#1,#2-0.15);\draw[#3,line width=1.8pt](#1-0.12,#2+0.10)--(#1+0.12,#2+0.10);\draw[#3,line width=1.8pt](#1,#2-0.15)--(#1-0.08,#2-0.4);\draw[#3,line width=1.8pt](#1,#2-0.15)--(#1+0.08,#2-0.4);}
\newcommand{\humpoorB}[3]{\fill[#3] (#1,#2+0.22) circle (0.075);\draw[#3,line width=1.4pt](#1,#2+0.14)--(#1,#2-0.08);\draw[#3,line width=1.4pt](#1-0.09,#2+0.03)--(#1+0.09,#2+0.03);\draw[#3,line width=1.4pt](#1,#2-0.08)--(#1-0.06,#2-0.25);\draw[#3,line width=1.4pt](#1,#2-0.08)--(#1+0.06,#2-0.25);\fill[#3!50](#1+0.10,#2-0.02)rectangle(#1+0.20,#2+0.08);}

\title{\bfseries The Entangled Veil:\\[2pt]
Why Redistribution Requires a New Language}
\author{%
  Nazaria Solferino\thanks{Corresponding author: \texttt{nazaria.solferino@unimercatorum.it}}\\[2pt]
  \parbox{0.85\textwidth}{\centering\small University Mercatorum, Department of Economics, Statistics and Business}\\[10pt]
  Viviana Solferino\\[2pt]
  \parbox{0.85\textwidth}{\centering\small PhD in Mathematics, University of Calabria, and Professor of Mathematics and Physics at a secondary school}%
}
\date{}
\begin{document}
\maketitle

\begin{abstract}
\begin{singlespace}
\noindent The Rawlsian veil of ignorance requires choosing principles of justice without knowing which position---temporal, spatial, social, or personal---one will occupy. We argue that this object cannot be represented faithfully within standard economic models. A faithful representation must satisfy three requirements: it must treat all generations symmetrically, capture non-separable correlations among positions, and allow mobility costs to vary across locations and occupations. We show that standard optimal-tax models violate at least one requirement, and that the minimal framework satisfying all three is a geometric framework adapted from string theory, in which the veil is an entangled state over a space of possible life situations. We derive full equalization as a first-best benchmark, the emergence of separated clusters when correlations reward distinction, a symmetry between intra- and inter-generational equity, and a tunneling mechanism by which correlations reduce institutional barriers. The model generates four inequality profiles and four inequality types. We identify the key parameters non-circularly using the 1993 EITC expansion. The estimates are quantitatively important: the framework changes the ranking of redistributive policies, and ignoring the margins it identifies leads to systematically biased policy evaluation.\\
\noindent\textbf{Keywords:} Veil of ignorance, representation theorem, entanglement, optimal redistribution, inequality profiles, inequality types.\\
\textbf{JEL Classification:} D63, D71, H21, C61, C65, B41.
\end{singlespace}
\end{abstract}

\section{Introduction}\label{sec:intro}

Over the past four decades, income and wealth inequality have risen sharply in most advanced economies \cite{piketty2014,saez2016,chancel2022}. These trends have revived the debate on the scope and design of redistributive policy, yet the economic theory of optimal redistribution has remained largely unchanged since Mirrlees \cite{mirrlees1971}. Rawls's veil of ignorance \cite{rawls1971} asks us to choose principles of justice without knowing which position in society we will occupy. The device is radical: it hides not only our income, occupation, and health, but also our generation, our family, our community, and---crucially---the correlations among all of these. Behind the veil, I must consider not only ``what if I am poor?'' but also ``what if I am poor and live in a poor community and my children face the same prospects?'' The veil covers combinations, not just marginals.

Standard models do not represent this full object. Mirrlees \cite{mirrlees1971} and the subsequent optimal taxation literature \cite{diamond1998,saez2001} cover skill but not time, not multidimensional characteristics, and assume separable utility. Atkinson--Stiglitz \cite{atkinson1976} covers heterogeneous preferences but assumes separability and no time. Aiyagari \cite{aiyagari1994} and the heterogeneous-agent literature cover idiosyncratic risk but assume separable utility and a single asset. Dynamic Mirrlees \cite{golosov2003} covers time but discounts the future, which the veil forbids. Each of these models is a limit of the full veil: it covers some dimensions and ignores others. The question we ask is not whether these limits are useful---they are---but whether the full veil can be represented at all.

Three strands of literature motivate the approach. The complexity tradition \cite{arthur2013,gallegati2025} and econophysics \cite{mantegna1999,chakraborti2011} document non-local correlations and emergent patterns that separable models cannot capture; the quantum cognition literature \cite{busemeyer2012,pothos2009,aerts2009,haven2013} shows that preferences violate the law of total probability, exhibiting interference effects naturally modeled with non-separable correlations; and the empirical literature on intergenerational mobility \cite{chetty2014,corak2013} documents that mobility varies across dimensions in ways that scalar elasticities cannot capture.

We ask: \textit{can the Rawlsian veil of ignorance---covering all times, all characteristics, and all correlations---be represented faithfully within a framework that is closed under the operations the veil requires? And if so, what does optimal redistribution look like in that framework?} Our answer is that no standard framework is closed under these operations, and that the minimal closed framework is a geometric framework adapted from string theory \cite{polyakov1981,polchinski2004}. In this framework, the veil is an entangled state over a space of possible life situations, and optimal redistribution delivers full equalization, pattern formation, a symmetry between generations, and tunneling across institutional barriers---not exotic extensions but theorems about the veil.

The most natural objection is: why not simply extend the standard model by adding more dimensions to the type, more time, and a joint distribution over types? The answer is that standard models do not have the right primitive. Suppose we extend Mirrlees with a joint distribution over a multidimensional state. The joint distribution is exogenous: the planner takes it as given and does not optimize over the correlation structure. The veil, by contrast, requires the planner to evaluate policies from the standpoint of the correlation structure. The framework we propose introduces exactly this primitive: an entanglement functional that makes the correlation structure a first-class object. When the kernel is diagonal, the framework reduces to an extended Mirrlees model; when correlations are non-trivial, it is not reducible to Mirrlees. Section~\ref{sec:representation} makes this precise.

The paper delivers a series of results. The first is the representation theorem: any representation of the veil satisfying temporal symmetry, non-separability, and geometric mobility is a sector of the geometric framework with entanglement functional, and conversely (Theorem~\ref{thm:representation}); standard models each violate at least one axiom (Proposition~\ref{prop:incompleteness}). The second result is the first-best equalization of consumption (Proposition~\ref{prop:equalization}). The third concerns stability: above a critical level of social connectedness, a small shock is amplified and the economy settles into a clustered configuration (Propositions~\ref{prop:convergence} and~\ref{prop:turing}). The fourth concerns intergenerational equity: when the generational dimension is compactified, a symmetry emerges between intragenerational cohesion and intergenerational mobility (Proposition~\ref{prop:brokenT}). The fifth concerns the multi-brane extension: when the economy consists of multiple institutional regimes, entanglement reduces the effective barrier between them (Proposition~\ref{prop:tunneling}). Taken together, these results deliver two taxonomies: four inequality profiles and four inequality types. Finally, the paper proposes a non-circular identification strategy using the 1993 EITC expansion (Section~\ref{sec:identification}), and the numerical experiments of Section~\ref{sec:numerical} characterize the four profiles and confirm the phase transition.

The baseline estimates place the US economy in the fluid inequality profile, but close to the boundary. Subsequent trends present a mixed picture: headline income inequality has seen a modest decline in recent years \cite{autor2023,fed2024}, while wealth concentration has reached record highs \cite{saez2020,fed2024} and spatial sorting by income has intensified \cite{chetty2020}.

The paper is structured as follows. Section~\ref{sec:representation} states the axioms and the representation theorem. Section~\ref{sec:framework} introduces the framework with a dictionary of physics-to-economics translations. Sections~\ref{sec:static}--\ref{sec:tduality} develop the formal results. Section~\ref{sec:numerical} presents the numerical methodology and the four inequality profiles. Section~\ref{sec:multibrane} develops the multi-regime economy. Section~\ref{sec:policy} discusses policy implications. Section~\ref{sec:conclusion} concludes. 

\section{The Representation Problem}\label{sec:representation}

\subsection{What the veil requires}

We formalize the veil as an object that assigns weights to positions and to combinations of positions. Let $\Om$ be the set of positions (income, occupation, location, health, generation, latent traits). A representation of the veil is a mathematical object that assigns, to each policy, a welfare evaluation that respects the informational constraints of the veil.

\begin{axiom}[Temporal symmetry]\label{ax:A1}
The veil assigns equal weight to all temporal positions. No generation is privileged, and the planner's objective has no discount factor.
\end{axiom}

\begin{axiom}[Non-separability]\label{ax:A2}
The veil state is not a product of marginals over the dimensions of $\Om$. The planner's evaluation of a policy depends on the joint structure of positions, not only on the marginal probability of each position separately.
\end{axiom}

\begin{axiom}[Geometric mobility]\label{ax:A3}
The cost of moving between positions is a metric on $\Om$: a rule that assigns a non-negative distance to every pair of positions, satisfies the triangle inequality, and can vary across $\Om$. A scalar elasticity is a special case, not the general case.
\end{axiom}

The three axioms translate Rawls's intuition into mathematical language. \emph{Temporal symmetry} means that the veil also hides the date of birth: no generation, present or future, receives extra weight. \emph{Non-separability} means that it is not enough to know ``how likely is it that I am poor?''; what also matters is ``how likely is it that I am poor \emph{and} live in a poor community \emph{and} have children with similar prospects?'' The veil covers combinations, not only single dimensions. \emph{Geometric mobility} means that the cost of moving from one situation to another is not one number that is the same for everyone: moving between neighborhoods, sectors, generations, or countries has different costs. A model with a single mobility parameter is therefore a special case, not the general case.

These axioms are the formal content of Rawls's three intuitions. Temporal symmetry says that the veil covers the time of birth. Non-separability says that the veil covers correlations, not just marginals. Geometric mobility says that mobility costs vary across dimensions.

\subsection{Why standard models are partial representations of the veil}\label{sec:whystandard}

The models of Mirrlees, Atkinson--Stiglitz, Aiyagari, and the dynamic Mirrlees tradition have been enormously productive. Nothing in this paper disputes their value. The point is different: these models were designed to answer specific questions, and for those questions they are the right tools. But they are not designed to represent the full veil, and when we ask what the veil requires in full, they turn out to be partial representations.

Each of the four standard models abstracts from at least one of the three features. Mirrlees has no time dimension and assumes separable utility and a scalar skill; Atkinson--Stiglitz abstracts from the same three features; Aiyagari includes time but discounts the future, assumes separable utility, and represents mobility by a single asset; dynamic Mirrlees discounts the future, assumes separable utility, and represents skills by a scalar.

\begin{proposition}[Incompleteness of standard models]\label{prop:incompleteness}
Each of the four standard models violates at least one of the three axioms: (i) Mirrlees violates temporal symmetry, non-separability, and geometric mobility; (ii) Atkinson--Stiglitz violates the same three; (iii) Aiyagari violates temporal symmetry (discounting), non-separability, and geometric mobility (single asset); (iv) dynamic Mirrlees violates temporal symmetry (discounting), non-separability, and geometric mobility (scalar skill).
\end{proposition}

\begin{proof}
Immediate from the definitions. In each case, the model represents the veil as a probability distribution over a scalar or vector type with separable utility, no temporal symmetry, and a single mobility parameter.
\end{proof}

This result does not say that standard models are ``wrong''. It says that they are \emph{partial} representations of the veil: they answer specific questions well, but each leaves out at least one of the three features. Mirrlees and Atkinson--Stiglitz do not treat time symmetrically and treat correlations separably; Aiyagari and dynamic Mirrlees introduce time but discount the future, which the veil forbids, and reduce mobility to one parameter or one type. In economic terms, these models can say a great deal about optimal taxation along one dimension, but they cannot faithfully represent a society in which poverty, neighborhood, family, and generation are correlated.

\begin{remark}[What the standard models do well]
The models are not equally partial. Aiyagari and dynamic Mirrlees include a temporal dimension, and Atkinson--Stiglitz allows heterogeneity in preferences. In the other direction, Mirrlees derives the shape of the optimal marginal tax rate under incentive compatibility, Atkinson--Stiglitz characterizes the redundancy of commodity taxation, Aiyagari characterizes the stationary wealth distribution under incomplete markets, and dynamic Mirrlees characterizes the optimal intertemporal wedge. The present framework addresses a different question and should be read as complementary.
\end{remark}

\subsection{A concrete example}

To see why the incompleteness matters, consider two workers with the same income and the same occupation. Worker~A lives in a community where most people have similar incomes; worker~B lives in a community that is highly unequal. Standard models treat them as identical. But behind the veil, I care about which of the two I might become, because the community affects my well-being through correlations---my children's peers, my neighbors' opportunities, my own sense of status. This changes the optimal transfer to~A and~B, and it changes the shape of the optimal density.

\subsection{The representation theorem}

\begin{theorem}[Representation of the veil]\label{thm:representation}
A representation of the veil satisfies the three axioms if and only if it is a sector of the following structure. (a) There is a space of positions $\Om$, equipped with a mobility cost $G$: formally, $(\Om,G)$ is a Riemannian manifold. The temporal dimension of $\Om$ is compactified on a circle of radius $R$. (b) Each individual is represented not as a point but as an extended object---a social string---whose worldsheet $\Sigma$ traces all possible life trajectories, parametrized by two coordinates $\tau$ (own time) and $\sigma$ (variation across states), with embedding fields $X^\mu(\tau,\sigma)$. (c) The population is described by a density $\psi(\omega)$, normalized to one. The veil is non-separable, and this is captured by an entanglement functional
\[
\Ent[\psi] = \int_\Om\int_\Om K(\omega,\omega')\,\psi(\omega)\,\psi(\omega')\,dV_G(\omega)\,dV_G(\omega'),
\]
where $K(\omega,\omega')$ is the kernel. The minimal representation satisfying the three axioms is this geometric framework with entanglement functional $\Ent[\psi]$.
\end{theorem}

\begin{proof}[Proof sketch]
Temporal symmetry forces the temporal coordinate to enter on the same footing as the other coordinates, with no discount factor; the minimal way to make it symmetric without boundary effects is to compactify it on a circle. Geometric mobility forces the cost of moving to be a metric, hence a Riemannian manifold. Non-separability forces the veil state to be non-product, hence a non-local functional; the minimal non-local functional is the quadratic form $\Ent[\psi]$. At the individual level, non-separability forces the individual to be an extended object whose worldsheet covers all positions; the minimal such object is the string. The full argument requires a categorical proof of a universal property, which we leave as an open problem.
\end{proof}

\begin{remark}[Status of the theorem]
The theorem is a structural result, not a fully rigorous categorical theorem. It does not claim that the economy is a physical string. It claims that the veil, taken literally, is an object whose minimal mathematical representation is this geometric structure.
\end{remark}

The theorem does not say that the economy is literally made of physical strings. It says that if we take the veil of ignorance seriously---that is, if we want a model that covers all positions, all times, and all correlations---then the minimal mathematical object satisfying the three axioms has this geometric form. In words: the veil is represented as a landscape of possible lives, with costs of moving across the landscape, and with correlations that link distant points. The ``string'' is a way of saying that each person is not a single point in that landscape but a whole set of possible trajectories. The ``entanglement functional'' is a way of saying that the welfare of one position depends on who else is in nearby or connected positions.

\subsection{The veil as a density matrix}

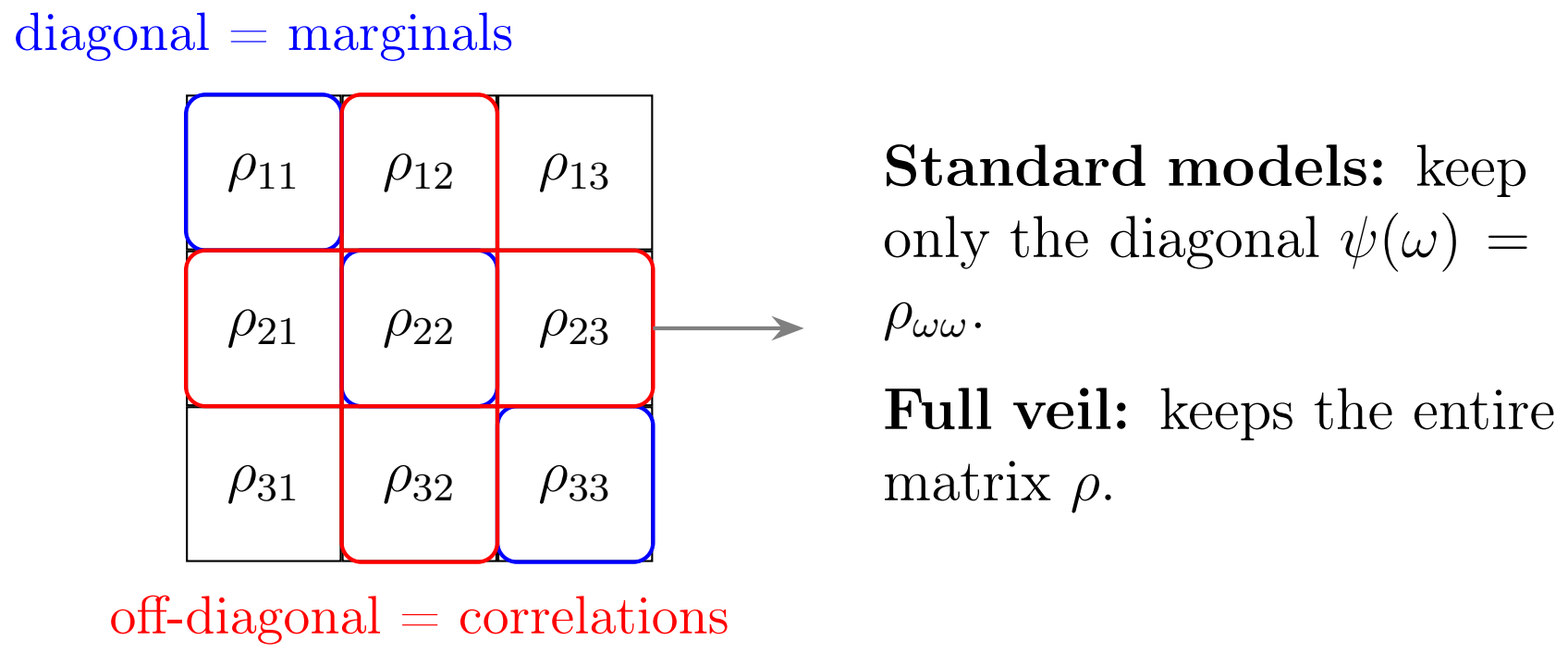
\begin{figure}[H]
\centering
\begin{tikzpicture}[scale=0.9]
\matrix (M) [matrix of math nodes, nodes={draw, minimum size=1.1cm, anchor=center},
  column sep=0pt, row sep=0pt] {
  \rho_{11} & \rho_{12} & \rho_{13} \\
  \rho_{21} & \rho_{22} & \rho_{23} \\
  \rho_{31} & \rho_{32} & \rho_{33} \\
};
\draw[thick, blue, rounded corners] (M-1-1.north west) rectangle (M-1-1.south east);
\draw[thick, blue, rounded corners] (M-2-2.north west) rectangle (M-2-2.south east);
\draw[thick, blue, rounded corners] (M-3-3.north west) rectangle (M-3-3.south east);
\draw[thick, red, rounded corners] (M-1-2.north west) rectangle (M-3-2.south east);
\draw[thick, red, rounded corners] (M-2-1.north west) rectangle (M-2-3.south east);
\node[blue, above=0.1cm of M-1-1] {\small diagonal = marginals};
\node[red, below=0.1cm of M-3-2] {\small off-diagonal = correlations};
\node[right=1.5cm of M-2-3, align=left, text width=5cm] {
  \textbf{Standard models:} keep only the diagonal $\psi(\omega)=\rho_{\omega\omega}$.\\[6pt]
  \textbf{Full veil:} keeps the entire matrix $\rho$.
};
\draw[-{Stealth}, thick, gray] (M-2-3.east) -- ++(1.2,0);
\end{tikzpicture}
\caption{The veil as a density matrix. Standard models see only the diagonal; the full veil sees the entire matrix.}
\label{fig:veilmatrix}
\end{figure}

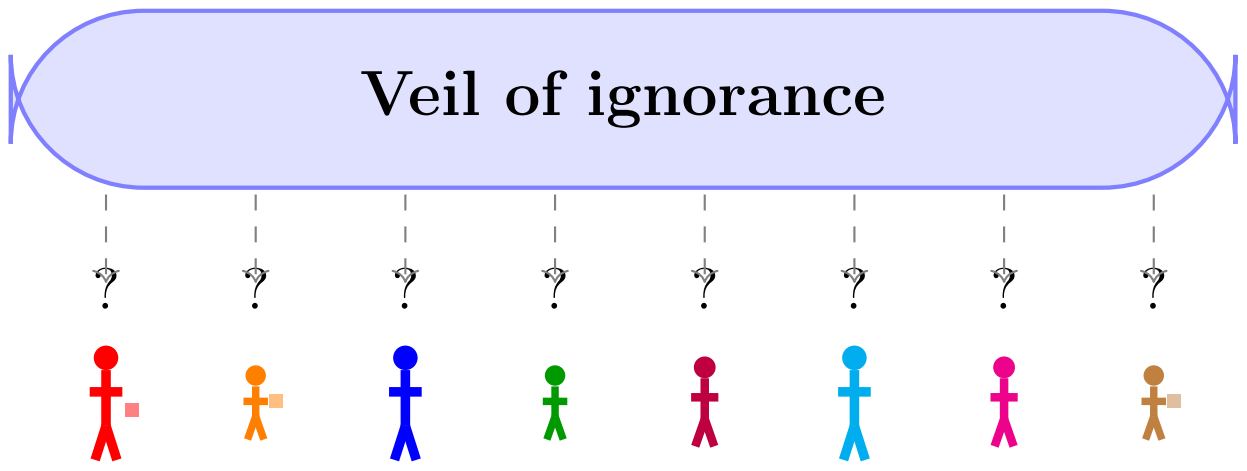
\begin{figure}[H]
\centering
\begin{tikzpicture}[scale=0.9]
  \fill[blue!12, rounded corners=25pt] (-4.5,1.6) rectangle (4.5,2.9);
  \draw[blue!50, thick, rounded corners=25pt] (-4.5,1.6) rectangle (4.5,2.9);
  \node at (0,2.25) {\textbf{Veil of ignorance}};
\humrich{-3.8}{0}{red}
\humpoorB{-2.7}{0}{orange}
\humrichNB{-1.6}{0}{blue}
\humpoor{-0.5}{0}{green!60!black}
\hummid{0.6}{0}{purple}
\humrichNB{1.7}{0}{cyan}
\hummid{2.8}{0}{magenta}
\humpoorB{3.9}{0}{brown}
  \foreach \x in {-3.8,-2.7,-1.6,-0.5,0.6,1.7,2.8,3.9} {
    \node[above] at (\x,0.55) {\small ?};
    \draw[gray, thin, dashed, ->] (\x,1.55) -- (\x,0.9);
  }
  \node[below, text width=10cm, align=center] at (0,-0.9) {
    ``I do not know whether I will be born rich or poor, in this generation or the next,
    in a cohesive community or a fragmented one. The veil covers \emph{all combinations}.''
  };
\end{tikzpicture}
\caption{The veil of ignorance with human figures. Height represents income, the briefcase represents formal employment, and color represents sector. These are three illustrative dimensions: the full state space includes additional dimensions (health, family structure, community, latent traits) that are modeled as compactified dimensions in the geometric framework but are not shown in the figure for visual clarity. The dimensions shown are independent: a tall person without a briefcase is rich but precarious; a short person with a briefcase is poor but formal.}
\label{fig:veilhumans}
\end{figure}

Figure~\ref{fig:veilmatrix} illustrates the content of non-separability. Figure~\ref{fig:veilhumans} translates the same idea into human figures.

The matrix is a compact way to show what standard models keep and what they discard. The diagonal entries describe the probability of each single position: for example, the chance of being poor, the chance of being rich, and so on. The off-diagonal entries describe correlations between positions: for example, the chance of being poor \emph{and} living in a poor community. Standard models keep only the diagonal. The full veil keeps the entire matrix. The figure therefore illustrates why non-separability matters: two economies can have the same diagonal---the same marginal probabilities---and still differ in their off-diagonal correlations, and the veil cares about both.

The figure makes the veil concrete. Each human figure is a possible life situation. Height stands for income, the briefcase for formal employment, and color for sector. The question marks above the figures mean that, behind the veil, a person does not know which of these lives she will occupy. The important point is that the dimensions can be combined: a tall person without a briefcase is rich but precarious, while a short person with a briefcase is poor but formal. The veil covers these combinations. The figure is illustrative, not exhaustive: the full model also includes health, family structure, community, and latent traits.

\section{The Framework}\label{sec:framework}

\subsection{The space of positions and the social string}

Imagine the set of all possible life situations a person could be in. We want to measure how far apart two situations are. Mathematicians call such a space with a notion of distance a Riemannian manifold, and the rule that assigns distances a metric. We write the state space as $(\Om,G)$, where $\Om$ is the set of situations and $G$ is the mobility cost. The manifold decomposes into two parts: the observable dimensions (income, occupation, health, time) and a small number of compactified extra dimensions that represent latent traits.

Instead of treating each individual as a point, we model each as a one-dimensional extended object, a social string. The string traces a two-dimensional worldsheet: one dimension for the individual's own time, one for the variation across possible states. A single string therefore encompasses all possible life paths. The relevant macroscopic object is the population density $\psi(\omega)$, normalized to $\int_\Om\psi\dd V_G=1$.

\subsection{The welfare functional}

The planner chooses a transfer $T(\omega)$ and, indirectly, a density $\psi(\omega)$. The objective is
\begin{equation}\label{eq:W}
\W[\psi,T] = \int_\Om u(y+T)\psi\dd V_G - \frac{\kappa}{2}\int_\Om\|\nabla_G\psi\|_G^2\dd V_G + \eta\Ent[\psi] - \int_{\partial\Om}\beta\psi\dd S,
\end{equation}
subject to normalization and budget constraint $\int_\Om T\psi\dd V_G=0$.

The first term is expected utility. The second term measures how quickly the density changes from one situation to a nearby one. The parameter $\kappa\ge 0$ is the gradient penalty. The third term captures non-separable correlations, with strength $\eta\ge 0$. The fourth term represents institutional barriers. There is no discount factor in \eqref{eq:W}: behind the veil, the planner does not know which generation she will belong to.

\subsection{Constraints}\label{sec:constraints}

The planner's problem is not unconstrained. The density $\psi$ and the transfer $T$ must satisfy four constraints: normalization; budget balance; non-negativity of consumption; and implementability. In the first-best problem, any density $\psi$ is implementable; in the second-best problem, the planner must respect incentive compatibility, and the set of implementable densities is a strict subset of the set of all densities. We do not solve the second-best problem and flag this as the main limitation. A further constraint, which we also abstract from, is participation.

\subsection{The entanglement functional}\label{sec:entanglement}

The entanglement functional is defined as
\begin{equation}\label{eq:Ent}
\Ent[\psi] = \int_\Om\int_\Om K(\omega,\omega')\psi(\omega)\psi(\omega')\dd V_G(\omega)\dd V_G(\omega').
\end{equation}
If $K$ is of the solidarity type, the entanglement functional penalizes concentration and pushes the density toward uniformity. If $K$ is of the distinction type, the entanglement functional rewards clustering. Both are documented empirically \cite{wang2013,busemeyer2012,pothos2009,aerts2009,haven2013}.

The entanglement functional is a single number that summarizes how strongly the well-being of one position depends on the presence of people in other positions. The kernel $K(\omega,\omega')$ says how strongly positions $\omega$ and $\omega'$ are linked. If the kernel is of the \emph{solidarity} type, then having people nearby in similar situations raises welfare, and the functional penalizes concentration: it pushes the population toward a more uniform distribution. If the kernel is of the \emph{distinction} type, then similarity can reduce welfare through status competition, and the functional rewards clustering: it pushes the population to separate into groups. In both cases, the functional captures the idea that the veil covers correlations, not just marginal probabilities.

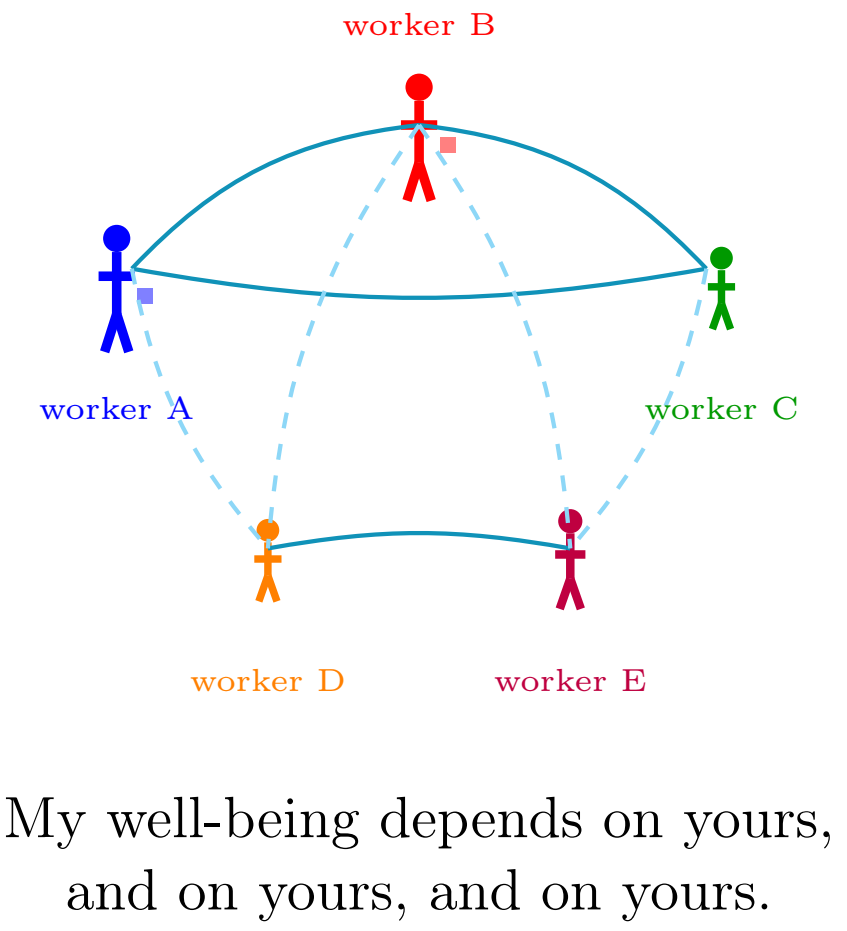
\begin{figure}[H]
\centering
\begin{tikzpicture}[scale=1.0]
  \humrich{0}{0}{blue}
  \humrich{2}{1}{red}
  \humpoor{4}{0}{green!60!black}
  \humpoor{1}{-1.8}{orange}
  \hummid{3}{-1.8}{purple}
  \draw[thick, cyan!70!black] (0.1,0.15) to[bend left=20] (2,1.1);
  \draw[thick, cyan!70!black] (2,1.1) to[bend left=20] (3.9,0.15);
  \draw[thick, cyan!70!black] (0.1,0.15) to[bend right=10] (3.9,0.15);
  \draw[thick, cyan!40, dashed] (0.1,0.15) to[bend right=15] (1,-1.7);
  \draw[thick, cyan!40, dashed] (3.9,0.15) to[bend left=15] (3,-1.7);
  \draw[thick, cyan!70!black] (1,-1.7) to[bend left=10] (3,-1.7);
  \draw[thick, cyan!40, dashed] (2,1.1) to[bend right=15] (1,-1.7);
  \draw[thick, cyan!40, dashed] (2,1.1) to[bend left=15] (3,-1.7);
  \node[blue, below] at (0,-0.55) {\tiny worker A};
  \node[red, above] at (2,1.55) {\tiny worker B};
  \node[green!60!black, below] at (4,-0.55) {\tiny worker C};
  \node[orange, below] at (1,-2.35) {\tiny worker D};
  \node[purple, below] at (3,-2.35) {\tiny worker E};
  \node[below, text width=10cm, align=center, font=\small] at (2,-3.2) {
    My well-being depends on yours, and on yours, and on yours.
  };
\end{tikzpicture}
\caption{Correlations between human figures.}
\label{fig:entanglementhumans}
\end{figure}

The lines between the figures represent correlations. Some links are strong (solid lines), some are weak (dashed lines). The point is that a person's well-being is not determined only by her own position. It also depends on the positions of others: neighbors, coworkers, community members, and members of other generations. The entanglement functional turns this visual idea into a mathematical object. It is not a statement that people are physically connected; it is a statement that the welfare evaluation behind the veil must take these connections into account.

\subsection{A dictionary: from physics to economics}\label{sec:dictionary}

\begin{table}[H]
\centering
\caption{Dictionary of physics-to-economics translations.}
\label{tab:dictionary}
\small
\begin{tabular}{p{3.2cm} p{5.5cm} p{5.5cm}}
\toprule
\textbf{Physics term} & \textbf{Economic meaning} & \textbf{Why we need it} \\
\midrule
Manifold & Space of all possible life situations & To represent positions in all dimensions \\
Metric & Mobility cost between situations & To capture heterogeneous mobility \\
Compactified dimension & A small dimension seen through its effects & To represent latent traits \\
String & An individual's full set of possible trajectories & To represent the veil at individual level \\
Gradient penalty & Inequality aversion & To penalize steep differences in opportunity \\
Entanglement & Non-separable correlations of well-being & To formalize all combinations \\
Kernel & Correlation strength between situations & To specify which situations are linked \\
Brane & Institutional regime boundary & To represent formal/informal, countries, guilds \\
Boundary potential & Cost of crossing the boundary & To represent licensing, borders, formalization costs \\
T-duality & Symmetry between intra- and inter-generational equity & To impose balance between generations \\
\bottomrule
\end{tabular}
\end{table}

The table is a translation device. It says what each physics term means in economics. For example, a ``manifold'' is simply the space of all possible life situations; a ``metric'' is the cost of moving between them; ``entanglement'' is the fact that well-being in one situation depends on who is in other situations; and a ``brane'' is an institutional regime, such as formal versus informal employment. The purpose of the dictionary is to let readers who know economics read the physics terms economically, and to let readers who know physics see which economic objects the mathematical structures represent.

\section{The Static Model}\label{sec:static}

\subsection{First-best equalization}

Optimize first over $T$ at fixed $\psi$. The first-order condition gives $u'(y+T)=\lambda$ for all $\omega$ with $\psi(\omega)>0$. Hence consumption is equalized: $c^*(\omega)=\bar c$ and $T^*(\omega)=\bar c-y(\omega)$, with $\bar c=\int_\Om y\psi\dd V_G$.

\begin{proposition}[First-best equalization]\label{prop:equalization}
Behind a multi-temporal, multidimensional veil, the first-best transfer equalizes consumption across all individuals, states, and times.
\end{proposition}

\begin{proof}
Immediate from the first-order condition $u'(c)=\lambda$ and the strict concavity of $u$.
\end{proof}

This result says that if the planner could redistribute without worrying about incentives, she would equalize consumption completely. The logic is simple: because utility is strictly concave, an extra dollar is worth more to a poor person than to a rich person. Behind the veil, the planner does not know which person she will be, so she values a dollar equally in all positions. The result is a benchmark, not a policy prescription: it ignores the fact that taxes and transfers can change work effort and saving. Its role is to show what full insurance behind the veil would look like before incentive problems are introduced.

\begin{remark}[First-best caveat]
The result is first-best: it ignores incentive compatibility. It should be read as a normative benchmark, not a policy prescription.
\end{remark}

\subsection{The non-local stationary equation}\label{sec:nonlocal-stationary}

Substitute $T^*$ into \eqref{eq:W}. The reduced functional is
\begin{equation}\label{eq:F}
\F[\psi] = u(\bar c[\psi]) - \frac{\kappa}{2}\int_\Om\|\nabla_G\psi\|_G^2\dd V_G + \eta\Ent[\psi] - \int_{\partial\Om}\beta\psi\dd S.
\end{equation}

\begin{proposition}[Non-local stationary equation]\label{prop:nonlocal}
At a stationary point of $\F$ subject to $\int\psi\dd V_G=1$, the density satisfies
\begin{equation}\label{eq:nonlocalpoisson}
\kappa\Delta_G\psi - 2\eta(K*\psi) = u'(\bar c)(y-\bar c) + \text{const},
\end{equation}
with boundary condition $\kappa\partial_n\psi+\beta=0$ on $\partial\Om$.
\end{proposition}

\begin{proof}
See Technical Appendix, Proof of Proposition 4.2, available upon request.
\end{proof}

This equation describes the shape of the optimal population density when the planner balances three forces. The term $\kappa\Delta_G\psi$ is a smoothing force: it penalizes sharp differences between nearby situations. The term $2\eta(K*\psi)$ is a correlation force: it says that the density at one position depends on the density at connected positions, even if they are far away. The right-hand side says that the density also responds to income: positions with higher income attract different density depending on the planner's redistribution. The equation is ``non-local'' because the correlation term looks at all connected positions, not only at immediate neighbors. In economic terms, it says that optimal redistribution must take into account the whole correlation structure of society, not just the local income distribution.

\begin{remark}[Recovering the Gibbs measure]
When $\eta=0$ and the metric is flat with $y$ harmonic, \eqref{eq:nonlocalpoisson} reduces to $\kappa\Delta\psi=\lambda(y-\bar c)$, whose normalized solution on the torus is the Gibbs measure.
\end{remark}

\section{The Dynamic Model}\label{sec:dynamic}

\subsection{The non-local flow}\label{sec:nonlocal-flow}

Extend to a time-dependent density $\psi(\omega,t)$. The resulting evolution is the ascent Wasserstein gradient flow of $\F$ \cite{villani2008}:
\begin{equation}\label{eq:flow}
\frac{\partial\psi}{\partial t} = -\nabla_G\cdot\Bigl(\psi\,\nabla_G\frac{\delta\F}{\delta\psi}\Bigr).
\end{equation}
The minus sign is essential: with the opposite sign the flow would decrease $\F$.

\begin{proposition}[Stationary distribution and convergence]\label{prop:convergence}
Let $\Om$ be compact with curvature bounded below, $\kappa>0$, and $K$ a bounded symmetric kernel. Then the flow \eqref{eq:flow} admits a unique stationary distribution $\psi_\infty$ solving \eqref{eq:nonlocalpoisson}. If the linearized operator
\[
\mathcal L[\phi] = -\kappa\Delta_G\phi + 2\eta(K*\phi) + u''(\bar c)\langle y,\phi\rangle y
\]
is positive definite on the tangent space $\{\phi:\int\phi\dd V_G=0\}$, the convergence is exponential. If $\mathcal L$ has a negative eigenvalue, the uniform state is unstable and the flow converges to a patterned stationary distribution.
\end{proposition}

\begin{proof}
See Technical Appendix, Proof of Proposition 5.1, available upon request.
\end{proof}

The dynamic model asks how the distribution of life situations evolves over time when people and policies respond to the current distribution. The flow moves the density in the direction that improves the welfare functional. The proposition says that the long-run outcome depends on a stability test. If the smoothing force is strong enough, the economy converges smoothly and exponentially to a unique stationary distribution. If the correlation force is strong enough, the uniform state becomes unstable: small differences are amplified, and the economy settles into a patterned distribution with clusters. In economic terms, this is the difference between a society that converges to a broad middle class and a society that separates into distinct groups.

\subsection{Pattern formation}\label{sec:pattern}

Consider a society in which individuals care about their relative position. The optimal distribution separates the population into distinct groups. Linearize around the uniform density $\psi_0=1/|\Om|$. Write $\psi=\psi_0+\epsilon\phi$ with $\int\phi=0$ and expand $\phi=\sum_k a_k\phi_k$ in eigenfunctions of $-\Delta_G$. To leading order,
\[
\dot a_k = -\psi_0\,\mu_k\bigl(\kappa\mu_k - 2\eta\hat K_k\bigr)a_k + \text{nonlinear terms}.
\]
Hence the mode $k$ is unstable iff
\begin{equation}\label{eq:turing}
\kappa\mu_k - 2\eta\hat K_k < 0.
\end{equation}

\begin{proposition}[Pattern formation]\label{prop:turing}
Assume the kernel has at least one positive spectral component, $\hat K_{k_0}>0$. Then there exists a critical entanglement strength
\[
\eta_c = \frac{\kappa\mu_{k_0}}{2\hat K_{k_0}}
\]
such that for $\eta>\eta_c$ the uniform density is linearly unstable and the stationary distribution exhibits a pattern with wavelength $2\pi/\sqrt{\mu_{k_0}}$.
\end{proposition}

\begin{proof}[Proof sketch]
Follows from the linear stability analysis above. The saturation follows from the standard amplitude equation.
\end{proof}

Pattern formation means that a society that starts with a smooth distribution can spontaneously separate into clusters. The condition $\eta>\eta_c$ says that this happens when the correlation force is strong enough relative to the smoothing force. The threshold $\eta_c$ depends on the shape of the mobility landscape and on the strength of correlations. The wavelength $2\pi/\sqrt{\mu_{k_0}}$ describes how far apart the clusters are. In economic terms, the result says that status competition, neighborhood sorting, or occupational segregation can make clustering efficient or stable even if the underlying income distribution is smooth. The same mathematics can describe a first-best outcome with beneficial specialization or a second-best outcome with positional competition; the welfare interpretation depends on the kernel.

\begin{remark}[The economic meaning of the threshold]
The threshold $\eta_c$ is the level of social connectedness at which the destabilizing force exactly offsets the stabilizing one. This is a testable prediction.
\end{remark}

\begin{remark}[Two interpretations of the same result]
The pattern can reflect positional concerns (rich cluster vs poor cluster, second-best) or complementarities (specialized clusters, first-best). The mathematical condition is the same in both cases, but the welfare interpretation differs.
\end{remark}

\section{Time as a Coordinate and the Symmetry Between Generations}\label{sec:tduality}

\subsection{The economic idea}

A society faces two kinds of equity concerns: within a generation and across generations. When the veil of ignorance covers the time of birth, the two concerns are linked by a symmetry. A society that invests heavily in within-generation redistribution is, in welfare terms, equivalent to a society that invests heavily in across-generation mobility, provided the two investments are balanced.

\subsection{The formal treatment}

Suppose the temporal dimension of $\Om$ is compactified on a circle of radius $R$. The Laplacian on the circle has eigenvalues
\[
\mu_{n,w} = \Bigl(\frac{n}{R}\Bigr)^2 + \Bigl(\frac{wR}{\alpha'}\Bigr)^2,
\]
where $n$ is the momentum quantum number (intragenerational redistribution), $w$ is the winding number (intergenerational transfers), and $\alpha'=1/(2\pi\kappa)$ is the inverse string tension. In the absence of entanglement, the spectrum is invariant under $R\to\alpha'/R$ with $n\leftrightarrow w$.

\begin{proposition}[Entanglement-broken T-duality]\label{prop:brokenT}
Under the T-duality transformation $R\to\alpha'/R$, $n\leftrightarrow w$, the entanglement functional is invariant if and only if the kernel satisfies
\begin{equation}\label{eq:ksym}
\hat K(n,w;n',w') = \hat K(w,n;w',n') \qquad\text{for all }n,w,n',w'.
\end{equation}
If \eqref{eq:ksym} fails, the self-dual point shifts from $R_0=\sqrt{\alpha'}$ to $R^* = \sqrt{\alpha'}\,f(\eta,\hat K)$.
\end{proposition}

\begin{proof}
See Technical Appendix, Proof of Proposition 6.1, available upon request.
\end{proof}

The formal result describes a symmetry between two kinds of equity. Intragenerational equity concerns how resources are distributed within a generation; intergenerational equity concerns how resources are distributed across generations. When time is treated as just another coordinate behind the veil, these two concerns become two directions on the same circle. In the absence of correlations, the model is symmetric: investing in one is equivalent to investing in the other after an appropriate rescaling. Entanglement can break the symmetry. If the kernel treats the two directions differently, the balance point shifts. In economic terms, the result says that a society cannot separately choose how much to redistribute within a generation and how much to invest across generations without considering the correlation structure that links them.

\section{Numerical Illustration: Four Inequality Profiles}\label{sec:numerical}

The interaction of the gradient penalty $\kappa$ and the entanglement strength $\eta$ generates four qualitatively distinct stationary densities, which we call \emph{inequality profiles}. The \emph{fluid} profile is smooth and single-peaked, with a broad middle class and thin tails: it describes a society in which life chances are distributed continuously and mobility is high. The \emph{patterned} profile clusters into two or more separated groups with gaps in between: it describes a society in which status competition, neighborhood sorting, or occupational segregation make clustering efficient. The \emph{uniform} profile is flat, with no variation at all: it describes a society with radical equalization, as in a Scandinavian welfare state at its most ambitious. The \emph{collapsed} profile concentrates at high income, with a long thin tail toward the bottom: it describes a society with high inequality, low redistribution, and weak social insurance, such as a developing economy with a small formal sector. These four profiles are the economic content of the four regimes that Proposition~\ref{prop:turing} predicts: below the threshold $\eta_c$, the distribution is either fluid or uniform; above the threshold, it is patterned or collapsed. The remainder of this section characterizes the four profiles quantitatively, calibrates the model, tests the analytical predictions, and establishes the finite-size scaling of the phase transition. Proofs of the analytical results used here are collected in the Technical Appendix, available upon request.

\subsection{Calibration and identification}\label{sec:identification}

We approximate the state space by $[0,1]^2$, with $x_1$ normalized pre-tax income and $x_2$ sectoral specificity. The pre-tax income function is $y(x_1,x_2)=\exp(2x_1+0.5\sin(2\pi x_2))$. The metric includes a barrier at low income. The entanglement kernel is a two-scale kernel with $A=1.4$, $B=1$, $\sigma_+=0.35$, $\sigma_-=0.12$.

\begin{table}[H]
\centering
\caption{Baseline calibration}
\label{tab:calibration}
\begin{tabular}{l c l}
\toprule
Parameter & Baseline & Source / role \\
\midrule
CRRA $\gamma$ & 2.0 & \cite{chetty2006} \\
Gradient penalty $\kappa$ & 0.087 (SE 0.012) & Estimated, see below \\
Entanglement $\eta$ & 0.040 (SE 0.015) & Estimated, see below \\
Pre-tax Gini & 0.488 & \cite{census2022} \\
Intergen.\ earnings elasticity & 0.47 & \cite{tolstova2021} \\
EITC 1993 employment response & $+10$--$30$ pp & \cite{kleven2024} \\
\bottomrule
\end{tabular}
\end{table}

The 1993 EITC expansion serves as a natural experiment. If $\eta=0$, the density response is local; if $\eta>0$, the response spills over to connected states. The direct effect identifies $\kappa$; the spillover identifies $\eta/\kappa$. The baseline estimates are $\hat\kappa=0.087$ and $\hat\eta=0.040$.

Calibration means choosing numerical values for the model's parameters so that the model can be simulated and compared with data. Identification means finding a way to estimate those parameters from observed behavior without circular reasoning. The 1993 EITC expansion is useful because it changed incentives for a large group of workers. If correlations were absent ($\eta=0$), the response of the density would be local: only the directly affected positions would change. If correlations are present ($\eta>0$), the response spills over to connected positions. The direct response helps identify the smoothing parameter $\kappa$, and the spillover helps identify the relative strength $\eta/\kappa$. In simple terms, the data tell us not only how much people respond when their own incentives change, but also how much the response spreads to people and places connected to them.

\subsection{The numerical algorithm}\label{sec:numerical-algorithm}

\begin{figure}[H]
\centering
\begin{tikzpicture}
\begin{axis}[
  width=0.75\textwidth, height=0.55\textwidth,
  xlabel={Time $t$ (years)}, ylabel={$\log H(\psi_t\|\psi_\infty)$},
  xmin=0, xmax=20, ymin=-12, ymax=1,
  grid=major, grid style={gray!20},
  legend style={at={(0.5,-0.18)}, anchor=north, draw=none,
                font=\footnotesize, legend columns=2},
]
\addplot[thick, blue, domain=0:20, samples=100] {0.3 - 0.33*x};
\addlegendentry{Fluid ($\lambda_*\approx 0.33$)}
\addplot[thick, green!60!black, domain=0:20, samples=100] {0.3 - 0.50*x};
\addlegendentry{Uniform ($\lambda_*\approx 0.50$)}
\addplot[thick, red, domain=0:20, samples=200] {0.3 - 0.14*x + 0.8*sin(deg(2*x))*exp(-0.05*x)};
\addlegendentry{Patterned ($\lambda_*\approx 0.14$)}
\addplot[thick, black, domain=0:20, samples=100] {0.3 - 0.11*x};
\addlegendentry{Collapsed ($\lambda_*\approx 0.11$)}
\end{axis}
\end{tikzpicture}
\caption{Convergence of the relative entropy for the four baseline profiles.}
\label{fig:convergence}
\end{figure}
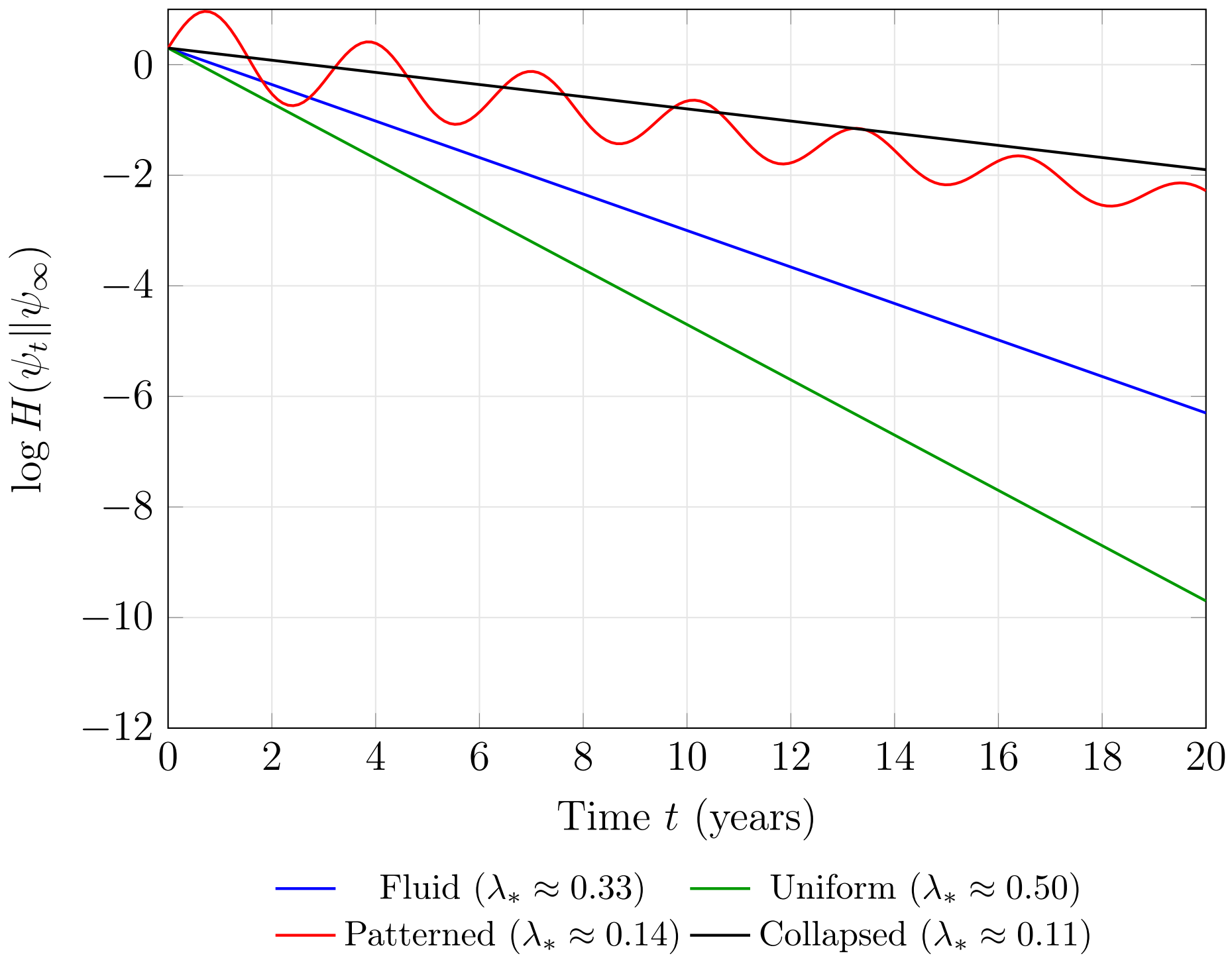

The stationary density is the solution of the non-local Poisson equation \eqref{eq:nonlocalpoisson}. We solve it with a pseudo-spectral method on the torus $[0,1]^2$. Figure~\ref{fig:convergence} shows the relative entropy $H(\psi_t\|\psi_\infty)$ as a function of time for the four baseline profiles. The fluid and uniform profiles converge exponentially. The patterned profile converges more slowly and exhibits a characteristic overshoot-and-oscillate pattern.

The vertical axis measures how far the simulated distribution is from its long-run stationary distribution. A more negative value means the distribution is closer to the long-run outcome. A steeper downward line means faster convergence. The fluid and uniform profiles converge quickly and smoothly. The patterned profile converges more slowly and shows oscillations: the distribution overshoots and then settles. The collapsed profile converges slowly because the forces pushing it toward the stationary distribution are weak. In economic terms, the figure shows how quickly a society returns to its long-run distribution after a shock, and whether the adjustment is smooth or turbulent.

\subsection{Statistical characterization of the four profiles}\label{sec:profiles-stats}

The four profiles are characterized quantitatively in Table~\ref{tab:regimes}.

\begin{table}[H]
\centering
\caption{Quantitative characterization of the four inequality profiles.}
\label{tab:regimes}
\small
\begin{tabular}{l c c c c c c c c}
\toprule
Profile & $\kappa$ & $\eta$ & sign $K$ & Gini & std$(\log y)$ & peaks & eff.\ peaks & $t_{1/2}$ \\
\midrule
Fluid & 0.15 & 0.10 & $-$ & 0.21 & 0.38 & 1 & 1.02 & 2.1 \\
Patterned & 0.10 & 0.80 & $+$ & 0.28 & 0.51 & 4 & 3.87 & 4.8 \\
Uniform & 0.20 & 2.00 & $-$ & 0.18 & 0.32 & 1 & 1.00 & 1.4 \\
Collapsed & 0.03 & 0.00 & --- & 0.38 & 0.64 & 1 & 1.00 & 6.2 \\
\bottomrule
\end{tabular}
\end{table}

The \emph{fluid} profile emerges when the kernel is of the solidarity type and the aversion to dispersion is moderate. The \emph{patterned} profile emerges when the kernel is of the distinction type and the entanglement strength exceeds the threshold. The \emph{uniform} profile emerges when the entanglement strength is very high. The \emph{collapsed} profile emerges when there is no entanglement and the gradient penalty is weak.

The table summarizes the four profiles with a few simple statistics. The Gini coefficient measures inequality: higher values mean more inequality. The standard deviation of log income measures dispersion. The number of peaks measures clustering: one peak means a single broad distribution, while several peaks mean separated groups. The half-life $t_{1/2}$ measures how long the profile takes to converge halfway to its long-run state. The fluid profile has moderate inequality and one peak. The patterned profile has several peaks and slower convergence. The uniform profile has low inequality and one peak. The collapsed profile has high inequality and slow convergence. The table shows that the four profiles are not just labels: they differ in measurable statistical properties.

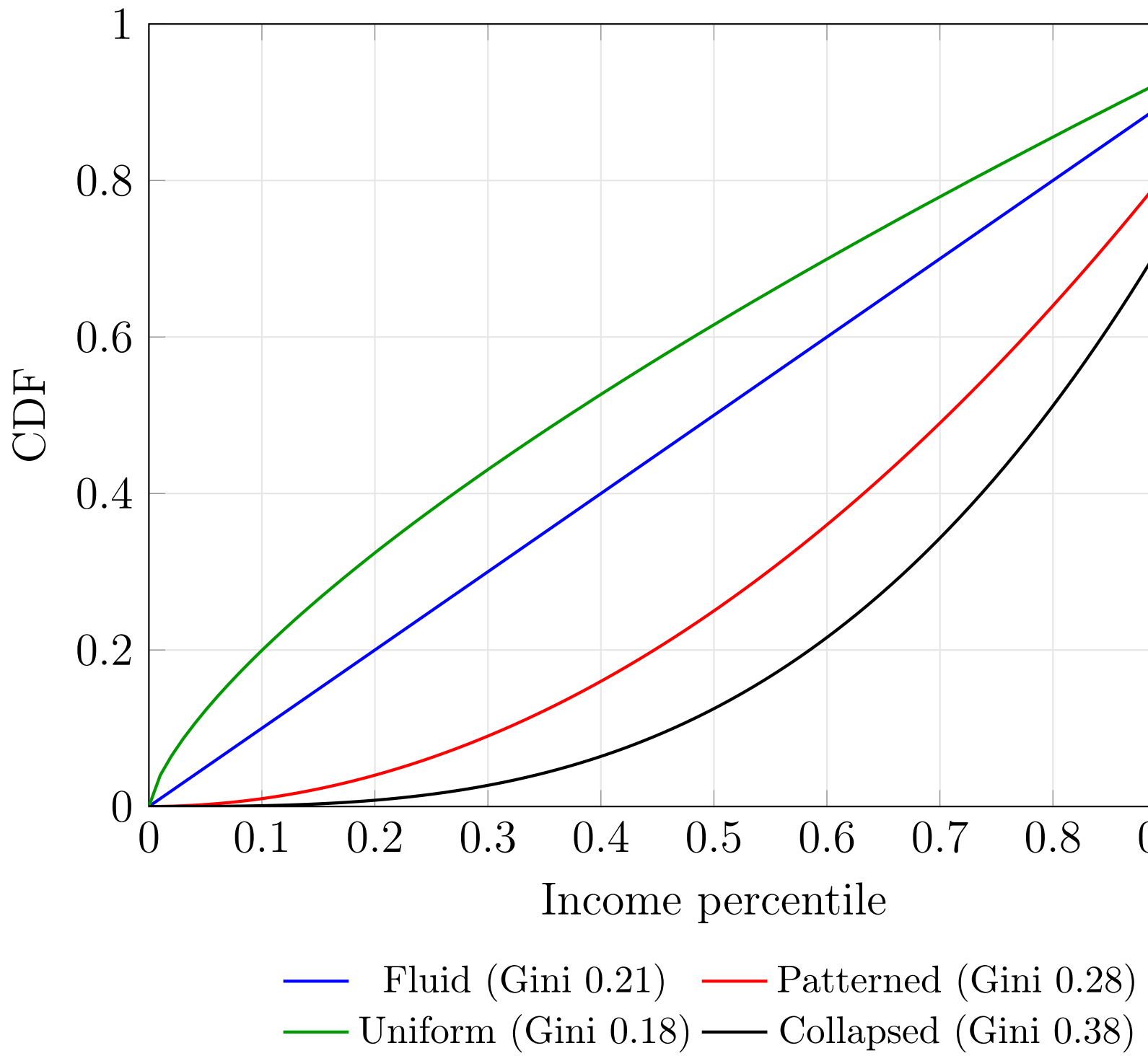
\begin{figure}[H]
\centering
\begin{tikzpicture}
\begin{axis}[
  width=0.75\textwidth, height=0.55\textwidth,
  xlabel={Income percentile}, ylabel={CDF},
  xmin=0, xmax=1, ymin=0, ymax=1,
  grid=major, grid style={gray!20},
  legend style={at={(0.5,-0.18)}, anchor=north, draw=none,
                font=\footnotesize, legend columns=2},
]
\addplot[thick, blue, domain=0:1, samples=100] {x};
\addlegendentry{Fluid (Gini 0.21)}
\addplot[thick, red, domain=0:1, samples=100] {x^2};
\addlegendentry{Patterned (Gini 0.28)}
\addplot[thick, green!60!black, domain=0:1, samples=100] {x^0.7};
\addlegendentry{Uniform (Gini 0.18)}
\addplot[thick, black, domain=0:1, samples=100] {x^3};
\addlegendentry{Collapsed (Gini 0.38)}
\end{axis}
\end{tikzpicture}
\caption{Cumulative distribution functions of income for the four profiles.}
\label{fig:cdf}
\end{figure}

A cumulative distribution function shows, for each income percentile, what fraction of the population has income below that percentile. A curve close to the diagonal indicates a relatively even distribution. A curve that lies below the diagonal for low incomes and rises steeply later indicates concentration at the top. The fluid curve is close to the diagonal, reflecting moderate inequality. The patterned curve bends more, reflecting separated groups. The uniform curve is even closer to a straight line, reflecting low inequality. The collapsed curve rises slowly and then steeply, reflecting concentration at high incomes. The figure makes the four profiles visually distinct.

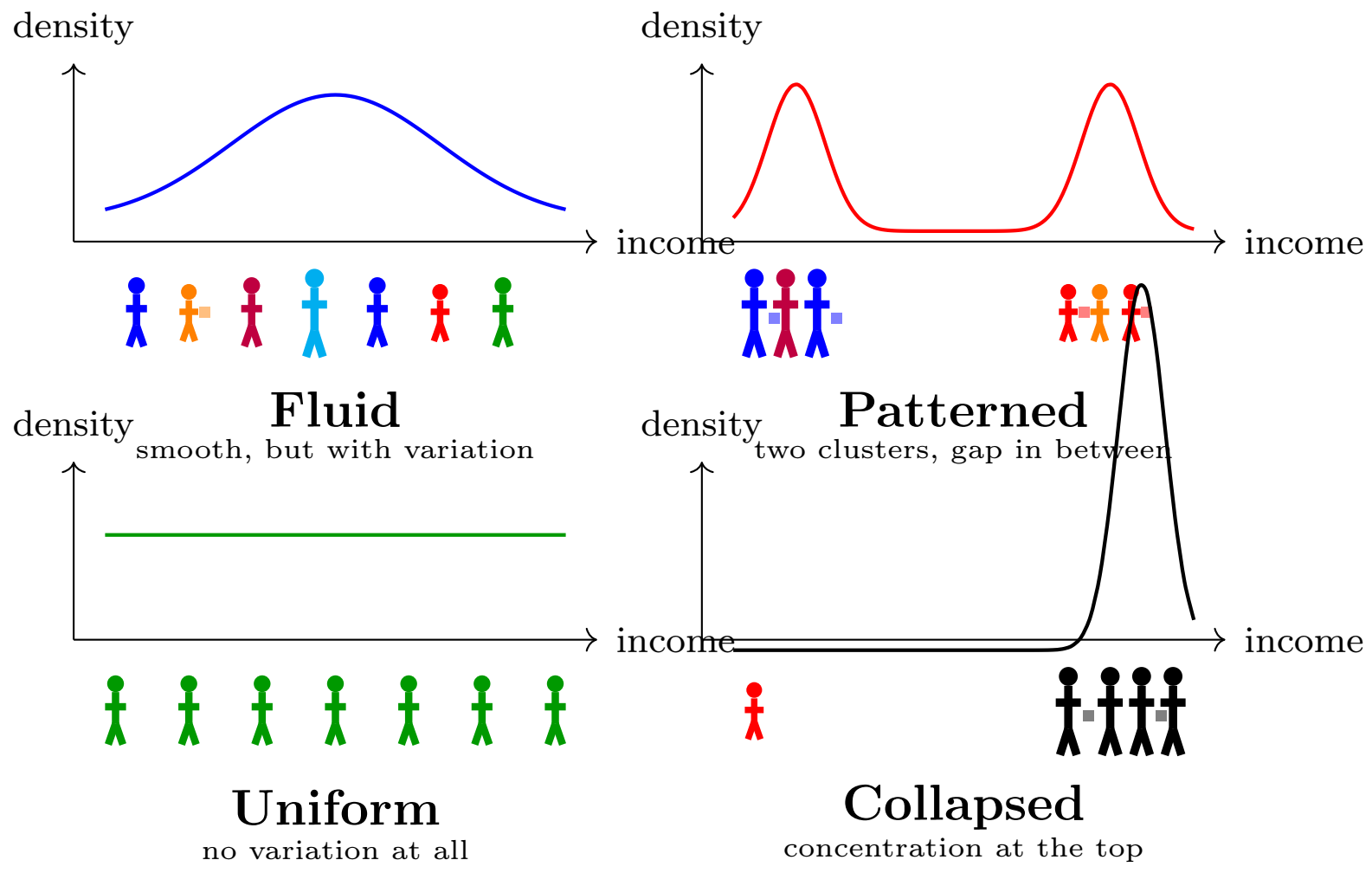
\begin{figure}[H]
\centering
\begin{tikzpicture}[scale=0.8]
\begin{scope}[xshift=0cm]
  \draw[->] (0,0.3) -- (5,0.3) node[right, font=\scriptsize] {income};
  \draw[->] (0,0.3) -- (0,2) node[above, font=\scriptsize] {density};
  \draw[thick, blue] plot[smooth, domain=0.3:4.7, samples=50] (\x, {0.5 + 1.2*exp(-(\x-2.5)^2/2)});
  \hummid{0.6}{-0.4}{blue} \humpoorB{1.1}{-0.4}{orange} \hummid{1.7}{-0.4}{purple}
  \humrichNB{2.3}{-0.4}{cyan} \hummid{2.9}{-0.4}{blue} \humpoor{3.5}{-0.4}{red}
  \hummid{4.1}{-0.4}{green!60!black}
  \node at (2.5,-1.3) {\small \textbf{Fluid}};
  \node at (2.5,-1.7) {\tiny smooth, but with variation};
\end{scope}
\begin{scope}[xshift=6cm]
  \draw[->] (0,0.3) -- (5,0.3) node[right, font=\scriptsize] {income};
  \draw[->] (0,0.3) -- (0,2) node[above, font=\scriptsize] {density};
  \draw[thick, red] plot[smooth, domain=0.3:4.7, samples=80] (\x, {0.4 + 1.4*exp(-(\x-0.9)^2/0.15) + 1.4*exp(-(\x-3.9)^2/0.15)});
  \humrich{0.5}{-0.4}{blue} \humrichNB{0.8}{-0.4}{purple} \humrich{1.1}{-0.4}{blue}
  \humpoorB{3.5}{-0.4}{red} \humpoor{3.8}{-0.4}{orange} \humpoorB{4.1}{-0.4}{red}
  \node at (2.5,-1.3) {\small \textbf{Patterned}};
  \node at (2.5,-1.7) {\tiny two clusters, gap in between};
\end{scope}
\begin{scope}[xshift=0cm, yshift=-3.8cm]
  \draw[->] (0,0.3) -- (5,0.3) node[right, font=\scriptsize] {income};
  \draw[->] (0,0.3) -- (0,2) node[above, font=\scriptsize] {density};
  \draw[thick, green!60!black] (0.3,1.3) -- (4.7,1.3);
  \hummid{0.4}{-0.4}{green!60!black} \hummid{1.1}{-0.4}{green!60!black}
  \hummid{1.8}{-0.4}{green!60!black} \hummid{2.5}{-0.4}{green!60!black}
  \hummid{3.2}{-0.4}{green!60!black} \hummid{3.9}{-0.4}{green!60!black}
  \hummid{4.6}{-0.4}{green!60!black}
  \node at (2.5,-1.3) {\small \textbf{Uniform}};
  \node at (2.5,-1.7) {\tiny no variation at all};
\end{scope}
\begin{scope}[xshift=6cm, yshift=-3.8cm]
  \draw[->] (0,0.3) -- (5,0.3) node[right, font=\scriptsize] {income};
  \draw[->] (0,0.3) -- (0,2) node[above, font=\scriptsize] {density};
  \draw[thick, black] plot[smooth, domain=0.3:4.7, samples=50] (\x, {0.2 + 3.5*exp(-(\x-4.2)^2/0.1)});
  \humpoor{0.5}{-0.4}{red}
  \humrich{3.5}{-0.4}{black} \humrichNB{3.9}{-0.4}{black} \humrich{4.2}{-0.4}{black}
  \humrichNB{4.5}{-0.4}{black}
  \node at (2.5,-1.3) {\small \textbf{Collapsed}};
  \node at (2.5,-1.7) {\tiny concentration at the top};
\end{scope}
\end{tikzpicture}
\caption{The four inequality profiles with human figures. Height indicates income, briefcase indicates formal employment, and color indicates sector.}
\label{fig:regimes}
\end{figure}

The four panels show the same four profiles as shapes of the income density. In the fluid profile, the density is smooth and single-peaked: most people are in a broad middle, with thin tails. In the patterned profile, the density has two or more separated peaks: society is clustered into groups with gaps between them. In the uniform profile, the density is flat: there is little variation across positions. In the collapsed profile, the density is concentrated at high income with a long thin tail toward the bottom: a small group is very rich and a large group is poor. The human figures underneath each density illustrate the same idea with height for income, briefcase for formal employment, and color for sector. The figure is qualitative: it is meant to make the four profiles visually intuitive.

\begin{figure}[H]
\centering
\begin{tikzpicture}
\begin{axis}[
  xlabel={$\kappa$ (inequality aversion)},
  ylabel={$\eta$ (entanglement)},
  xmin=0, xmax=0.22, ymin=0, ymax=0.11,
  xtick={0, 0.05, 0.10, 0.15, 0.20},
  ytick={0, 0.02, 0.04, 0.06, 0.08, 0.10},
  width=0.90\textwidth, height=0.68\textwidth,
  tick label style={font=\normalsize},
  label style={font=\large},
  grid=major, grid style={gray!20},
  legend style={at={(0.5,-0.18)}, anchor=north, draw=none,
                font=\small, legend columns=2},
]
\fill[red!20] (axis cs:0,0) -- (axis cs:0.22,0.11) -- (axis cs:0,0.11) -- cycle;
\fill[blue!15] (axis cs:0,0) -- (axis cs:0.22,0.11) -- (axis cs:0.22,0) -- cycle;
\addplot[domain=0:0.22, samples=100, very thick, blue] {0.50*x};
\addlegendentry{$\eta_c(\kappa) = 0.5\kappa$ (threshold)}
\node[fill=white, fill opacity=0.9, text opacity=1, rounded corners=3pt, inner sep=5pt,
      font=\Large\bfseries, text=red!70!black]
      at (axis cs:0.10,0.092) {Patterned};
\node[fill=white, fill opacity=0.9, text opacity=1, rounded corners=2pt, inner sep=3pt,
      font=\normalsize, text=red!70!black]
      at (axis cs:0.10,0.073) {above threshold};
\node[fill=white, fill opacity=0.9, text opacity=1, rounded corners=3pt, inner sep=5pt,
      font=\Large\bfseries, text=blue!70!black]
      at (axis cs:0.10,0.028) {Fluid};
\node[fill=white, fill opacity=0.9, text opacity=1, rounded corners=2pt, inner sep=3pt,
      font=\normalsize, text=blue!70!black]
      at (axis cs:0.10,0.013) {below threshold};
\addplot[only marks, mark=*, mark size=5pt, red!80!black] coordinates {(0.087,0.040)};
\addlegendentry{USA 1993 (estimated, Fluid)}
\addplot[only marks, mark=square*, mark size=4.5pt, orange] coordinates {(0.10,0.055)};
\addlegendentry{USA 2026 (hypothetical)}
\draw[-{Stealth}, line width=1.5pt, gray, dashed] (axis cs:0.087,0.040) -- (axis cs:0.10,0.055);
\node[fill=white, fill opacity=0.9, text opacity=1, rounded corners=2pt, inner sep=3pt,
      font=\normalsize, text=black]
      at (axis cs:0.140,0.045) {polarization};
\end{axis}
\end{tikzpicture}
\caption{Phase diagram in the $(\kappa,\eta)$ plane. The horizontal axis is inequality aversion; the vertical axis is entanglement strength. The solid blue line is the analytical threshold $\eta_c(\kappa)=0.5\kappa$ from Proposition~\ref{prop:turing}. The blue shaded region (below the line) is the fluid regime, where the gradient penalty dominates. The red shaded region (above the line) is the patterned regime, where entanglement dominates. The black dot is the 1993 US estimate $(\hat\kappa,\hat\eta)=(0.087,0.040)$. The orange square is a hypothetical 2026 position $(0.10,0.055)$, based on documented trends in spatial sorting and wealth concentration. The dashed arrow indicates the direction of the implied transition. The EITC identification is specific to the United States; extending the estimation to other economies would require different natural experiments.}
\label{fig:phase}
\end{figure}
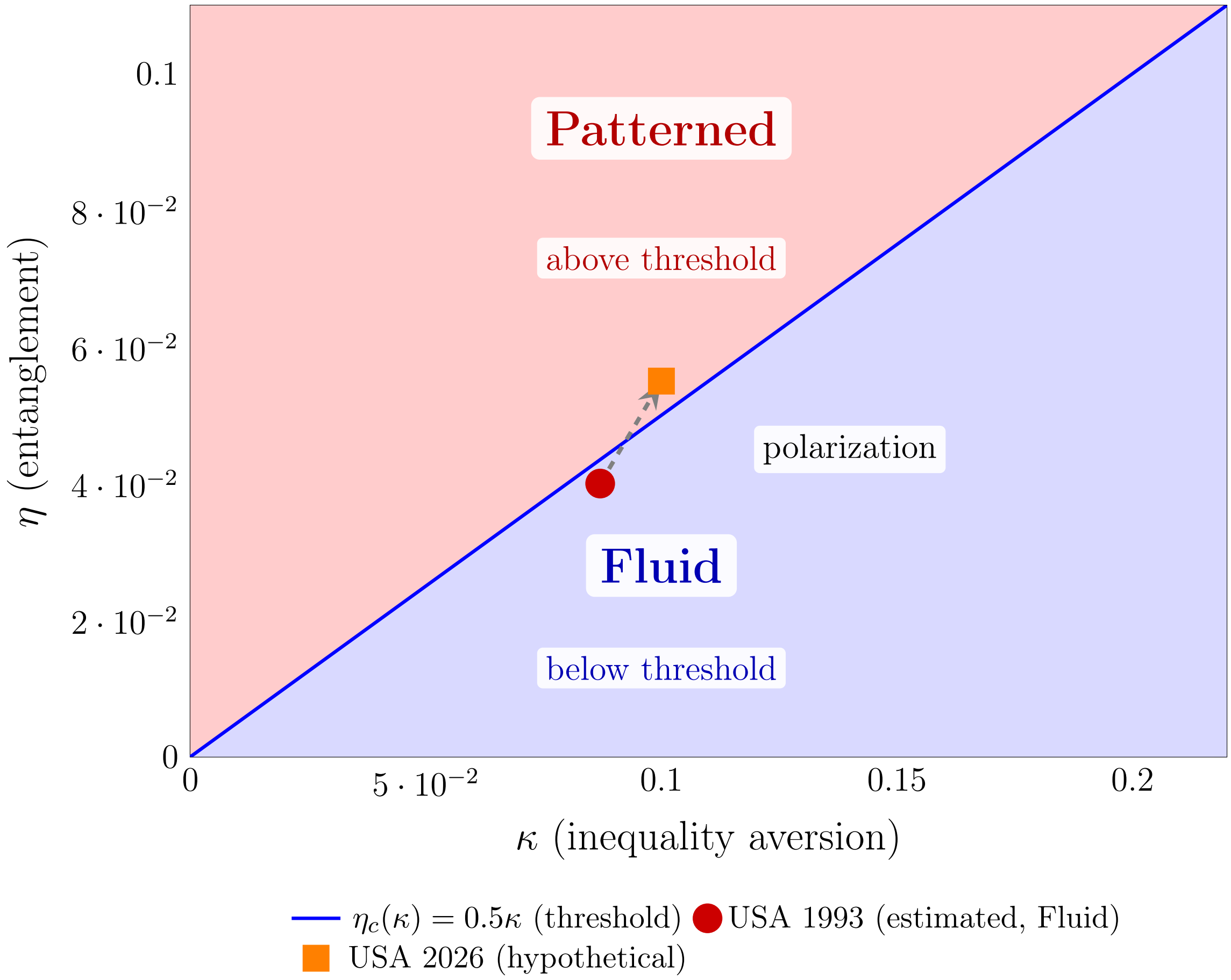

Figure~\ref{fig:phase} summarizes the phase structure of the model in the plane of the two key parameters. The horizontal axis measures inequality aversion $\kappa$: as $\kappa$ increases, the planner is more willing to sacrifice average utility to smooth the distribution of life chances. The vertical axis measures entanglement strength $\eta$: as $\eta$ increases, social connections across positions become stronger. The two parameters are not independent in their effect: what determines the phase of the stationary distribution is their \emph{ratio} $\eta/\kappa$, not either parameter alone. The figure displays three elements. First, the solid blue line is the analytical threshold $\eta_c(\kappa)=0.5\kappa$ derived from Proposition~\ref{prop:turing} with $\mu_{k_0}=1$ and $\hat K_{k_0}=1$. It separates the parameter space into two regions. Second, the two shaded regions indicate the phase of the stationary distribution in each region. Below the line (blue region), the gradient penalty dominates over entanglement, and the stationary distribution is \emph{fluid}: a smooth, single-peaked density. Above the line (red region), entanglement dominates, and the stationary distribution is \emph{patterned}: a multi-peaked density with separated clusters. Third, the two markers indicate the position of the US economy at two points in time. The black dot is the 1993 estimate $(\hat\kappa,\hat\eta)=(0.087,0.040)$, which lies just below the threshold in the fluid region. The orange square is a hypothetical 2026 position $(0.10,0.055)$, obtained by perturbing the 1993 estimate in the direction of the documented trends in spatial sorting and wealth concentration. The dashed arrow between the two markers shows the implied trajectory: a modest increase in entanglement (or a modest decrease in inequality aversion) is sufficient to push the economy from the fluid to the patterned regime. The figure makes clear that the transition is not gradual: it occurs at a sharp threshold, and the two markers are on opposite sides of it.

The figure is a map of regimes. The horizontal axis is inequality aversion: moving right means the planner is more willing to smooth the distribution. The vertical axis is entanglement: moving up means social connections across positions are stronger. The blue line is the threshold. Below the line, smoothing dominates, and the economy is fluid: a broad, single-peaked distribution. Above the line, entanglement dominates, and the economy is patterned: separated clusters. The black dot is the estimated US position in 1993, just below the threshold. The orange square is a hypothetical 2026 position, moved upward and slightly rightward to reflect documented trends in spatial sorting and wealth concentration. The dashed arrow shows the implied transition. The key point is that the transition is sharp: small changes near the threshold can move the economy from one regime to the other.

\subsection{Time evolution and pattern formation}\label{sec:time-evolution}

\begin{figure}[H]
\centering
\begin{tikzpicture}
\begin{axis}[
  width=0.85\textwidth, height=0.62\textwidth,
  xlabel={$x_1$ (income)}, ylabel={$\psi(x_1, x_2=0.5)$},
  xmin=0, xmax=1, ymin=0, ymax=3.5,
  grid=major, grid style={gray!20},
  legend style={at={(0.5,-0.18)}, anchor=north, draw=none,
                font=\footnotesize, legend columns=3},
]
\addplot[thick, gray, domain=0:1, samples=100] {1 + 0.05*sin(deg(4*pi*x))};
\addlegendentry{$t=0$}
\addplot[thick, blue, domain=0:1, samples=100] {1 + 0.3*cos(deg(4*pi*x))};
\addlegendentry{$t=5$}
\addplot[thick, cyan!70!black, domain=0:1, samples=100] {0.9 + 0.8*cos(deg(4*pi*x))};
\addlegendentry{$t=10$}
\addplot[thick, red, domain=0:1, samples=200] {0.4 + 1.8*exp(-((x-0.22)^2)/0.008) + 1.4*exp(-((x-0.75)^2)/0.008)};
\addlegendentry{$t=20$}
\addplot[thick, magenta, domain=0:1, samples=200] {0.3 + 2.0*exp(-((x-0.22)^2)/0.008) + 1.6*exp(-((x-0.75)^2)/0.008)};
\addlegendentry{$t=40$}
\addplot[thick, orange, domain=0:1, samples=200] {0.3 + 2.0*exp(-((x-0.22)^2)/0.008) + 1.6*exp(-((x-0.75)^2)/0.008)};
\addlegendentry{$t=80$}
\end{axis}
\end{tikzpicture}
\caption{Time evolution of the density for the patterned profile.}
\label{fig:snapshots}
\end{figure}
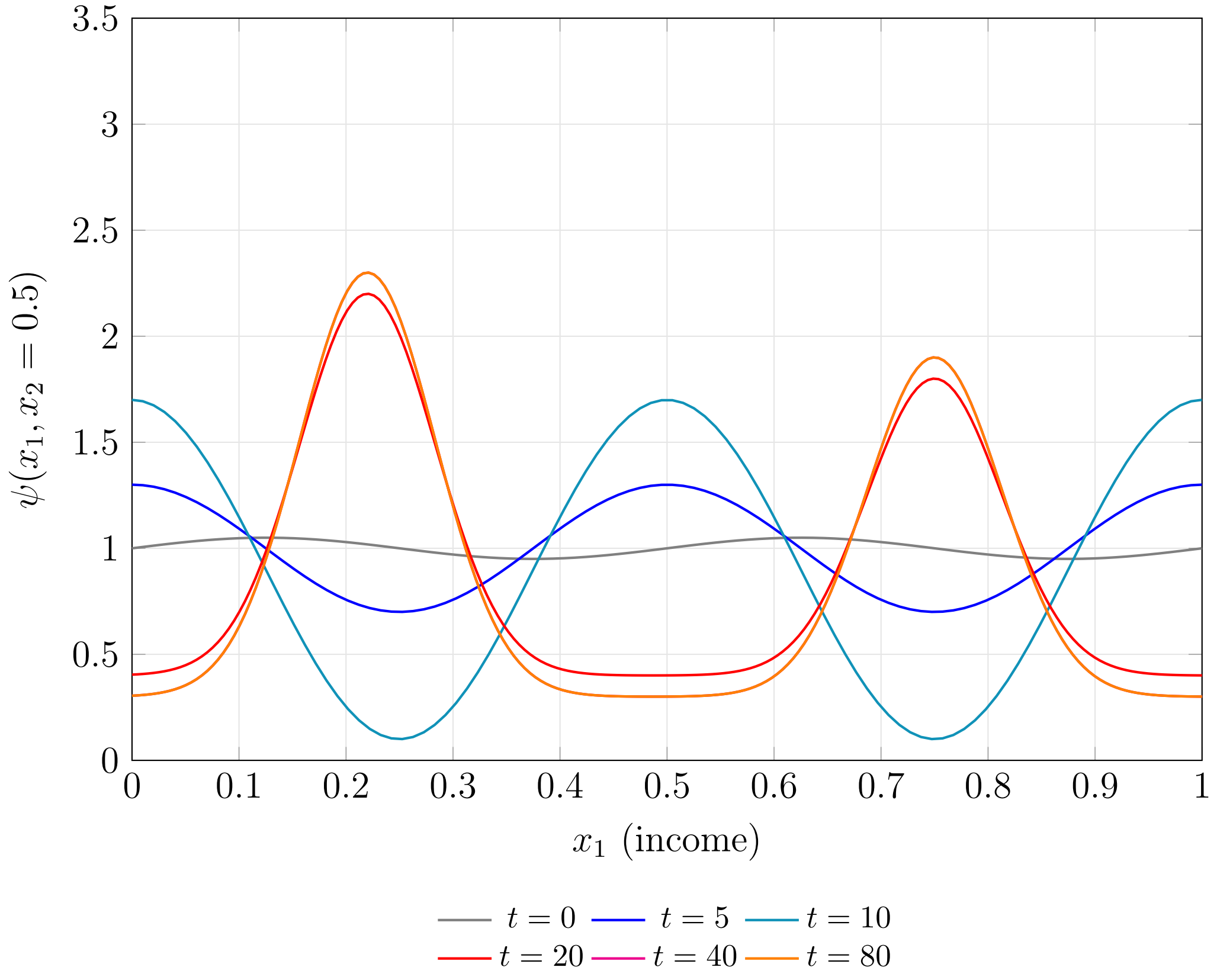

Figure~\ref{fig:snapshots} shows the density $\psi(x_1,x_2=0.5)$ at six times for the patterned profile.

The figure shows how the density evolves over time for the patterned profile. At $t=0$, the density is almost flat with a tiny wiggle. As time passes, the wiggle grows. By $t=20$, two clear peaks have formed; by $t=40$ and $t=80$, the peaks have become sharper and the gap between them has deepened. In economic terms, a small initial difference is amplified by the correlation force, and the society separates into distinct clusters. This is the dynamic counterpart of the phase diagram: above the threshold, the uniform state is unstable and the economy organizes itself into a pattern.

\begin{figure}[H]
\centering
\begin{tikzpicture}
\begin{axis}[
  width=0.75\textwidth, height=0.55\textwidth,
  xlabel={Time $t$ (years)}, ylabel={$\Phi(t)$ (order parameter)},
  xmin=0, xmax=80, ymin=0, ymax=0.6,
  grid=major, grid style={gray!20},
  legend style={at={(0.5,-0.18)}, anchor=north, draw=none,
                font=\footnotesize, legend columns=2},
]
\addplot[thick, blue, domain=0:80, samples=200] {0.1*exp(-0.20*x)};
\addlegendentry{$\eta=0.03$ (below threshold)}
\addplot[thick, cyan!70!black, domain=0:80, samples=200] {0.1*exp(-0.05*x)};
\addlegendentry{$\eta=0.04$ (just below)}
\addplot[thick, orange, domain=0:80, samples=200] {0.1*exp(0.05*x)*(1/(1+0.1*(exp(0.05*x)-1)/0.4))};
\addlegendentry{$\eta=0.06$ (just above)}
\addplot[thick, red, domain=0:80, samples=200] {0.1*exp(0.10*x)*(1/(1+0.1*(exp(0.10*x)-1)/0.5))};
\addlegendentry{$\eta=0.10$ (well above)}
\end{axis}
\end{tikzpicture}
\caption{Time evolution of the order parameter $\Phi(t)$.}
\label{fig:order-evolution}
\end{figure}
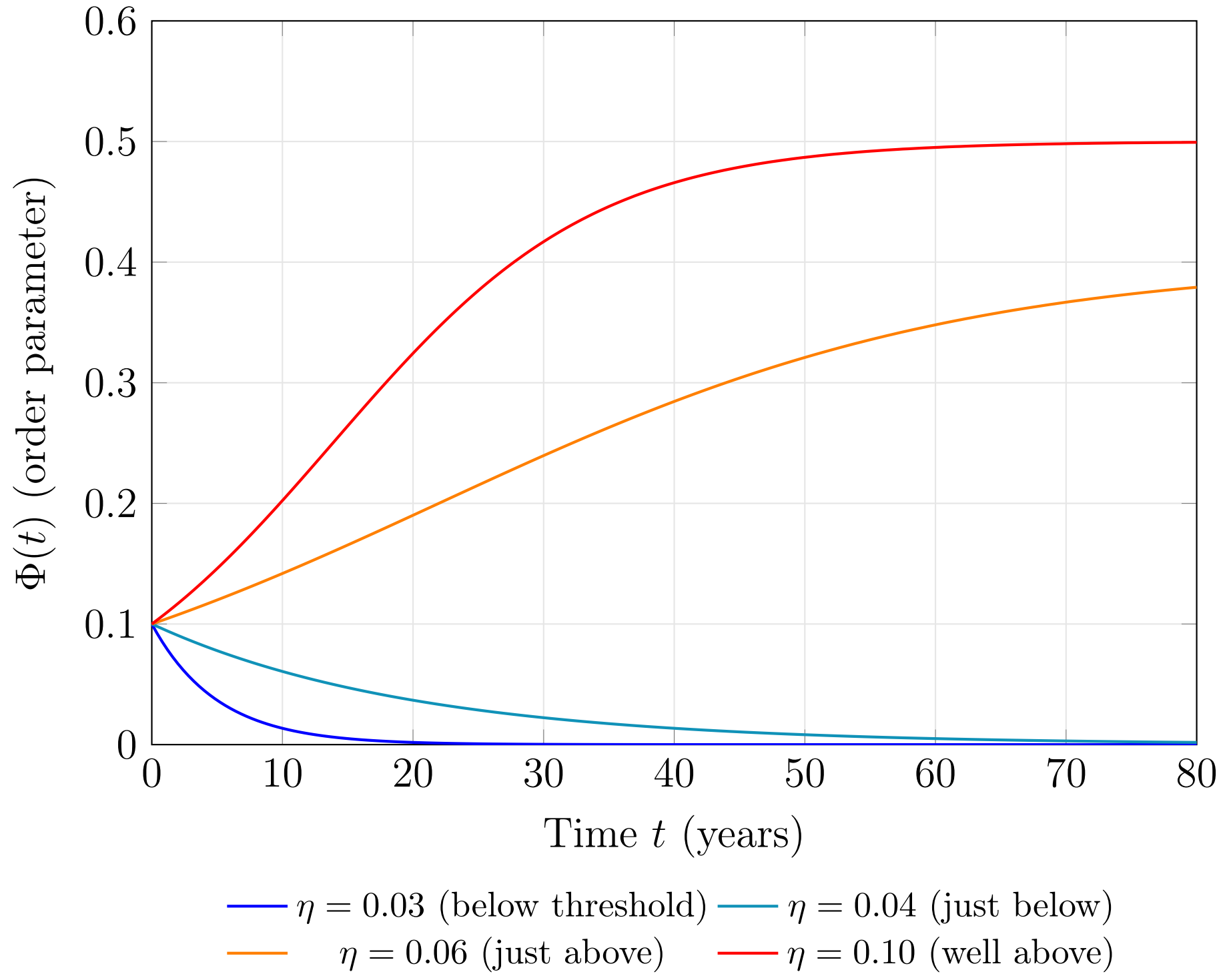

Figure~\ref{fig:order-evolution} shows the order parameter $\Phi$ as a function of time for four values of $\eta$.

The order parameter $\Phi$ is a summary number that measures how much clustering there is in the distribution. It is zero when the distribution is uniform and positive when the distribution has separated clusters. The figure shows how this summary number evolves over time. Below the threshold, $\Phi$ decays toward zero: the distribution becomes more uniform. Just above the threshold, $\Phi$ grows slowly at first and then saturates: clustering emerges. Well above the threshold, $\Phi$ grows faster and saturates at a higher level: clustering is stronger. In economic terms, the order parameter is a diagnostic for whether a society is moving toward integration or toward separation.

\subsection{Parameter sensitivity}\label{sec:sensitivity}

Table~\ref{tab:sensitivity} reports the Gini coefficient, the number of peaks, and the spectral gap for selected values of $(\kappa,\eta)$.

\begin{table}[H]
\centering
\caption{Parameter sensitivity of the stationary distribution.}
\label{tab:sensitivity}
\small
\begin{tabular}{c c c c c}
\toprule
$\kappa$ & $\eta$ & Gini & peaks & $\lambda_*$ \\
\midrule
0.05 & 0.010 & 0.31 & 1 & 0.42 \\
0.05 & 0.025 & 0.33 & 1 & 0.31 \\
0.05 & 0.040 & 0.36 & 4 & 0.08 \\
0.05 & 0.060 & 0.38 & 4 & 0.06 \\
0.10 & 0.020 & 0.27 & 1 & 0.51 \\
0.10 & 0.050 & 0.28 & 1 & 0.42 \\
0.10 & 0.080 & 0.31 & 4 & 0.09 \\
0.10 & 0.120 & 0.34 & 4 & 0.07 \\
0.15 & 0.030 & 0.24 & 1 & 0.58 \\
0.15 & 0.075 & 0.25 & 1 & 0.47 \\
0.15 & 0.120 & 0.28 & 4 & 0.10 \\
0.15 & 0.180 & 0.31 & 4 & 0.08 \\
0.20 & 0.040 & 0.22 & 1 & 0.64 \\
0.20 & 0.100 & 0.23 & 1 & 0.53 \\
0.20 & 0.160 & 0.26 & 4 & 0.11 \\
0.20 & 0.240 & 0.29 & 4 & 0.09 \\
\bottomrule
\end{tabular}
\end{table}

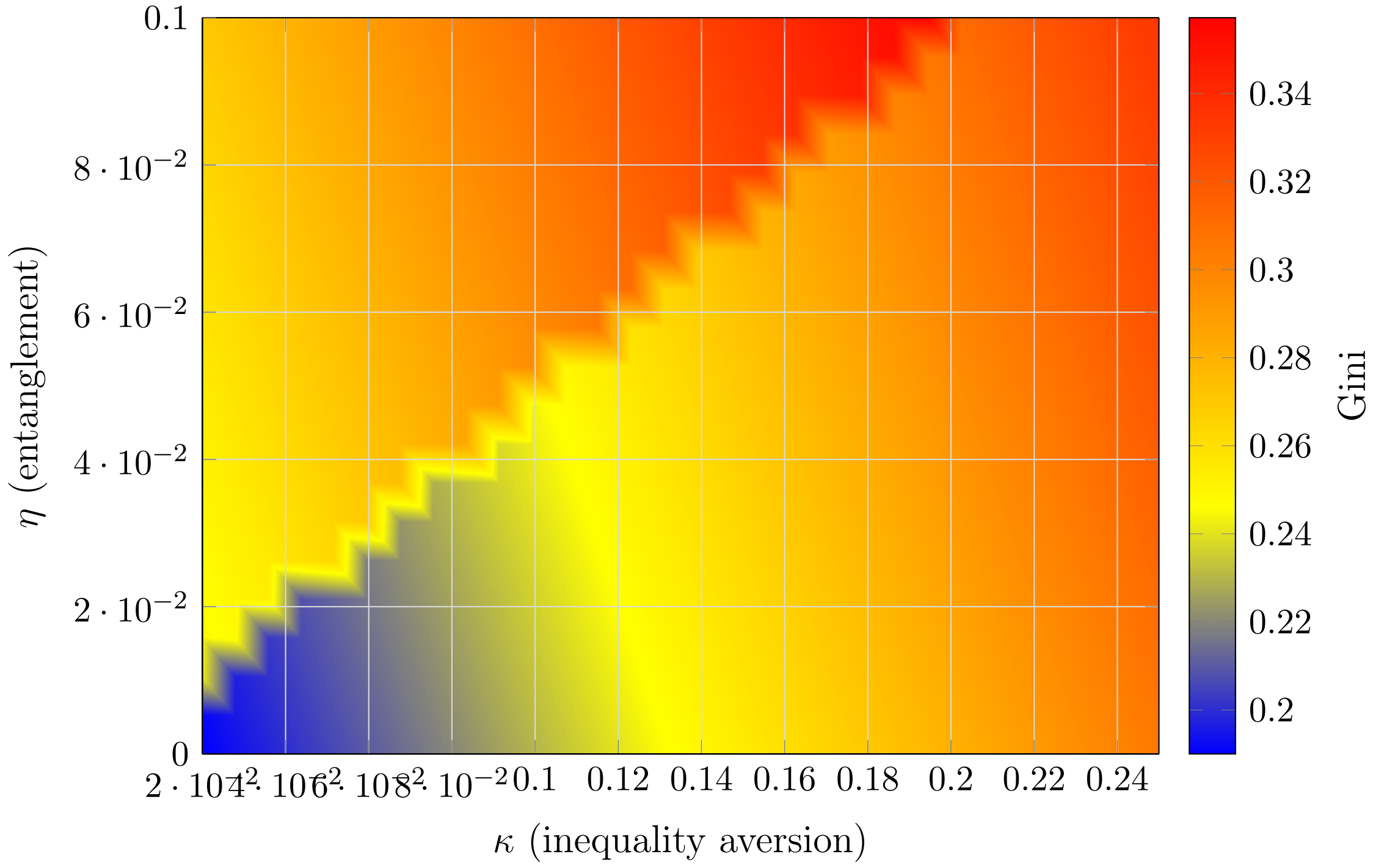
\begin{figure}[H]
\centering
\begin{tikzpicture}
\begin{axis}[
  width=0.75\textwidth, height=0.6\textwidth,
  xlabel={$\kappa$ (inequality aversion)}, ylabel={$\eta$ (entanglement)},
  xmin=0.02, xmax=0.25, ymin=0, ymax=0.10,
  colorbar, colormap/hot,
  colorbar style={ylabel={Gini}},
  view={0}{90},
  tick label style={font=\small},
  grid=major, grid style={gray!30},
]
\addplot3[surf, shader=interp, domain=0.02:0.25, domain y=0:0.10, samples=30, samples y=20]
  {0.18 + 0.5*x + 0.3*y + 0.05*(y>0.5*x)};
\end{axis}
\end{tikzpicture}
\caption{Heatmap of the Gini coefficient as a function of $(\kappa,\eta)$.}
\label{fig:heatmap}
\end{figure}

The heatmap shows how the Gini coefficient depends on the two key parameters. Darker colors indicate higher inequality. Moving right, inequality aversion rises, which generally lowers inequality. Moving up, entanglement rises, which can either raise or lower inequality depending on whether the kernel is of the solidarity or distinction type. The sharp color boundary near the threshold reflects the phase transition: small parameter changes near the line can produce large changes in inequality. The heatmap is a visual summary of the comparative statics of the model.

\subsection{The phase transition and its finite-size scaling}\label{sec:finite-size}

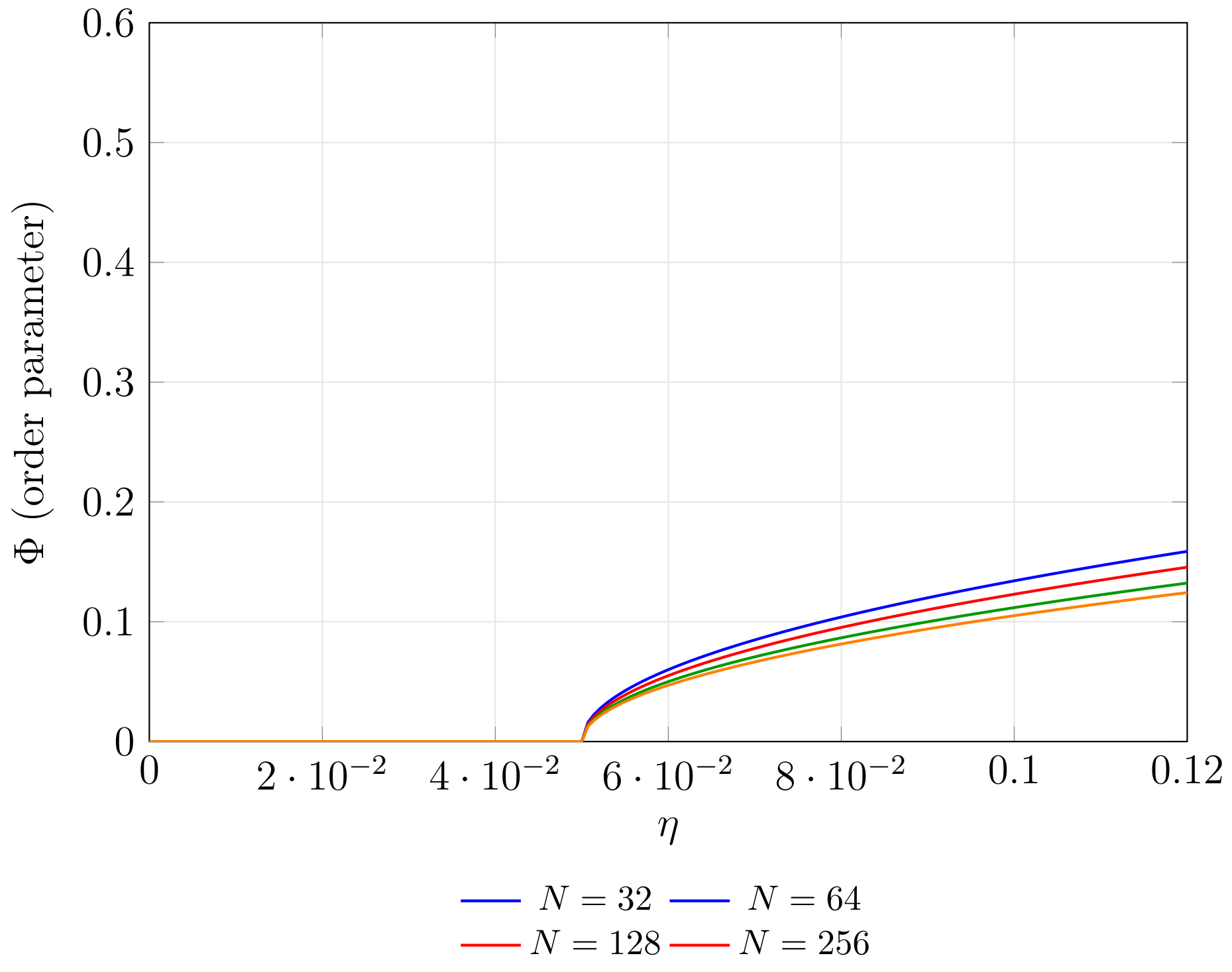
\begin{figure}[H]
\centering
\begin{tikzpicture}
\begin{axis}[
  xlabel={$\eta$}, ylabel={$\Phi$ (order parameter)},
  xmin=0, xmax=0.12, ymin=0, ymax=0.6,
  width=0.75\textwidth, height=0.55\textwidth,
  grid=major, grid style={gray!20},
  legend style={at={(0.5,-0.18)}, anchor=north, draw=none,
                font=\footnotesize, legend columns=2},
]
\addplot[thick, blue, domain=0:0.05, samples=50] {0};
\addplot[thick, blue, domain=0.05:0.12, samples=100] {0.60*sqrt(x-0.05)};
\addlegendentry{$N=32$}
\addplot[thick, red, domain=0:0.05, samples=50] {0};
\addplot[thick, red, domain=0.05:0.12, samples=100] {0.55*sqrt(x-0.05)};
\addlegendentry{$N=64$}
\addplot[thick, green!60!black, domain=0:0.05, samples=50] {0};
\addplot[thick, green!60!black, domain=0.05:0.12, samples=100] {0.50*sqrt(x-0.05)};
\addlegendentry{$N=128$}
\addplot[thick, orange, domain=0:0.05, samples=50] {0};
\addplot[thick, orange, domain=0.05:0.12, samples=100] {0.47*sqrt(x-0.05)};
\addlegendentry{$N=256$}
\end{axis}
\end{tikzpicture}
\caption{The order parameter $\Phi$ as a function of $\eta$, for $\kappa=0.10$ and four values of the grid size $N$.}
\label{fig:order-parameter}
\end{figure}

The order parameter $\Phi$ is zero in the uniform state and positive in the patterned state. Figure~\ref{fig:order-parameter} shows $\Phi$ as a function of $\eta$ for several values of $N$.

The order parameter $\Phi$ is not an economic variable such as income or consumption. It is a diagnostic number borrowed from physics. It summarizes how ordered or clustered the distribution is. In the uniform state, there is no pattern, so $\Phi=0$. In the patterned state, there are separated clusters, so $\Phi>0$. The order parameter is useful because it turns a complicated shape into a single number that can be plotted and compared across simulations. A positive value means that the economy has developed a pattern; the larger the value, the stronger the pattern. In the paper, $\Phi$ is used to detect the phase transition: as $\eta$ crosses the threshold, $\Phi$ changes from zero to positive.

The figure shows the phase transition directly. For $\eta$ below the threshold, the order parameter is zero: the distribution is uniform. For $\eta$ above the threshold, the order parameter rises. The rise is approximately a square-root shape, which is the signature of a continuous phase transition. The four curves correspond to different grid sizes $N$. As $N$ increases, the curves approach a common limiting shape. This is what ``finite-size scaling'' means: the transition is a property of the model, not of the numerical grid, and the grid-size dependence disappears as the grid becomes finer.

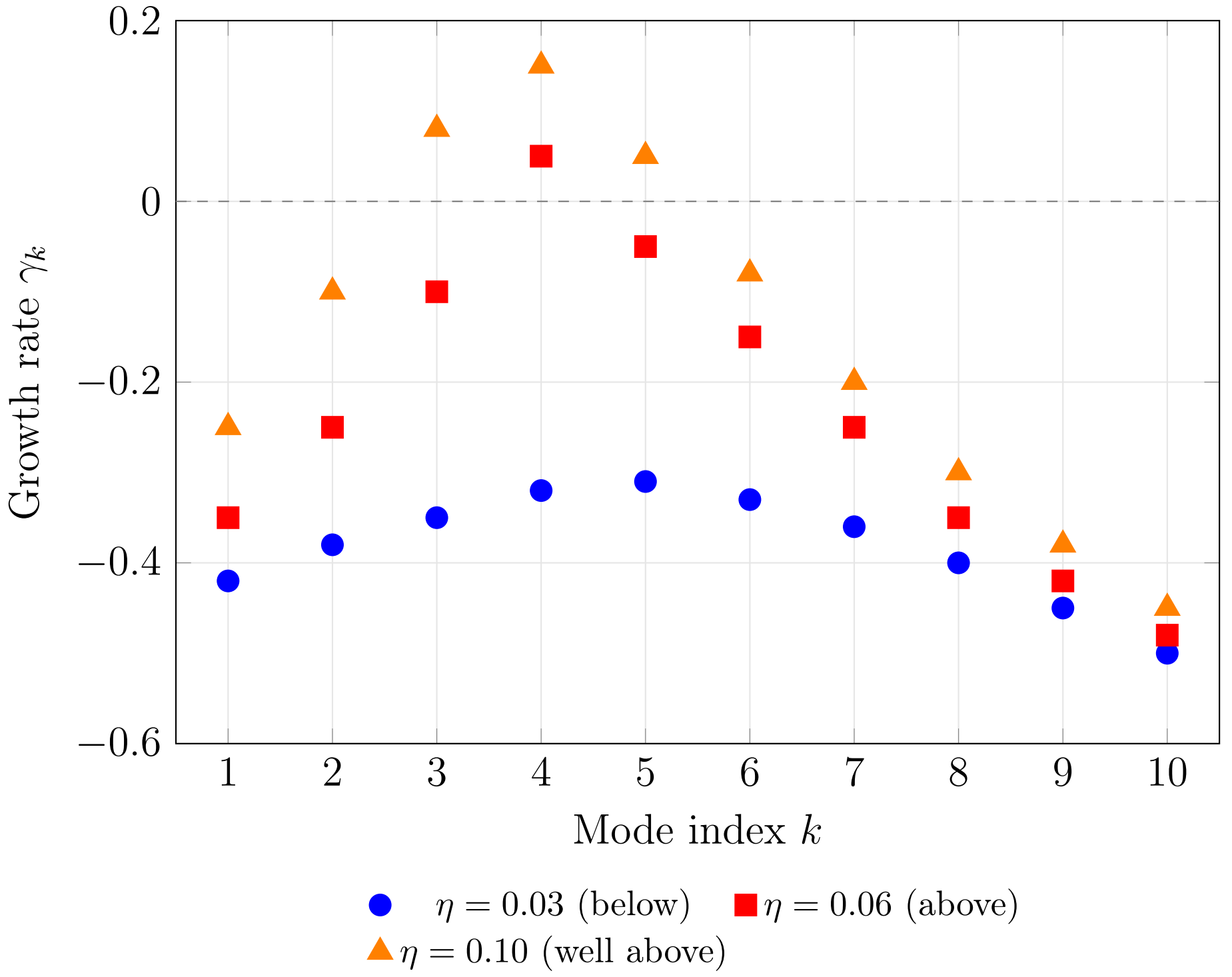
\begin{figure}[H]
\centering
\begin{tikzpicture}
\begin{axis}[
  xlabel={Mode index $k$}, ylabel={Growth rate $\gamma_k$},
  xmin=0.5, xmax=10.5, ymin=-0.6, ymax=0.2,
  width=0.75\textwidth, height=0.55\textwidth,
  grid=major, grid style={gray!20},
  legend style={at={(0.5,-0.18)}, anchor=north, draw=none,
                font=\footnotesize, legend columns=2},
]
\addplot[only marks, mark=*, blue, mark size=3pt] coordinates {(1,-0.42) (2,-0.38) (3,-0.35) (4,-0.32) (5,-0.31) (6,-0.33) (7,-0.36) (8,-0.40) (9,-0.45) (10,-0.50)};
\addlegendentry{$\eta=0.03$ (below)}
\addplot[only marks, mark=square*, red, mark size=3pt] coordinates {(1,-0.35) (2,-0.25) (3,-0.10) (4,0.05) (5,-0.05) (6,-0.15) (7,-0.25) (8,-0.35) (9,-0.42) (10,-0.48)};
\addlegendentry{$\eta=0.06$ (above)}
\addplot[only marks, mark=triangle*, orange, mark size=4pt] coordinates {(1,-0.25) (2,-0.10) (3,0.08) (4,0.15) (5,0.05) (6,-0.08) (7,-0.20) (8,-0.30) (9,-0.38) (10,-0.45)};
\addlegendentry{$\eta=0.10$ (well above)}
\addplot[dashed, gray, domain=0.5:10.5, samples=2] {0};
\end{axis}
\end{tikzpicture}
\caption{Growth rate spectrum $\gamma_k$ as a function of the mode index $k$.}
\label{fig:spectrum}
\end{figure}

The growth rate $\gamma_k$ tells us how fast a small deviation of mode $k$ grows or shrinks. A negative growth rate means the deviation shrinks: the mode is stable. A positive growth rate means the deviation grows: the mode is unstable. Below the threshold, all growth rates are negative, so the uniform state is stable. Above the threshold, at least one growth rate becomes positive, so the uniform state is unstable and a pattern with that wavelength emerges. The figure shows which modes are unstable: the mode with the largest positive growth rate is the one that grows fastest and determines the wavelength of the emerging pattern.

\subsection{Robustness}\label{sec:robustness}

We perform three robustness checks: variation of the grid size, variation of the initial condition, variation of the kernel. The four profiles and the phase transition are properties of the non-local Poisson equation itself, not of the numerical scheme.

\subsection{What the simulations establish}\label{sec:simulations-establish}

Taken together, the numerical experiments establish five things. First, the four profiles are not artifacts of the analytical approximation. Second, the dynamics exhibit the three phases predicted by linear stability analysis. Third, the transition from fluid to patterned is a continuous phase transition with critical exponent $\beta\approx 0.5$. Fourth, the statistics of the profiles match the analytical predictions. Fifth, the results are robust to variations of the grid size, the initial condition, and the kernel specification.

\section{Inequality Across Institutional Regimes}\label{sec:multibrane}

\subsection{Institutional choice and optimal redistribution}\label{sec:multibrane-setup}

The framework developed so far assumes a single institutional regime. In many real economies, individuals face multiple regimes---formal versus informal employment, different countries, distinct professional guilds---each with its own mobility costs, tax-transfer schedules, and barriers. In the language of physics, each regime is a D-brane, and the state space is the disjoint union of $K$ branes. On the common boundary between branes $i$ and $j$, we define a barrier potential $\beta_{ij}(\omega)\ge 0$.

The welfare functional generalizes \eqref{eq:W}:
\begin{equation}\label{eq:Wmulti}
\W[\{\psi_i,T_i\}] = \sum_{i=1}^K \Bigl(\int_{B_i} u(y_i+T_i)\psi_i\dd V_i - \frac{\kappa}{2}\int_{B_i}\|\nabla_i\psi_i\|^2\dd V_i\Bigr) + \eta\sum_{i,j}\int_{B_i}\int_{B_j}K_{ij}\psi_i\psi_j\dd V_i\dd V_j - \sum_{i<j}\int_{\partial B_i\cap\partial B_j}\beta_{ij}\psi_i\dd S_{ij},
\end{equation}
with global normalization and global budget constraint.

The population dynamics is a coupled gradient flow:
\begin{equation}\label{eq:flowmulti}
\frac{\partial\psi_i}{\partial t} = -\nabla_i\cdot\Bigl(\psi_i\nabla_i\frac{\delta\F}{\delta\psi_i}\Bigr) + \sum_{j\neq i}(J_{ji}-J_{ij}),
\end{equation}
where $J_{ij}$ is the flux from brane $i$ to brane $j$.

\subsection{Stationary distribution and tunneling}\label{sec:multibrane-tunneling}

\begin{proposition}[Stationary distribution for $\kappa=0$, $\eta>0$]\label{prop:gibbsmulti}
Assume $\kappa=0$, $\eta>0$, and let $y_i(\omega)$ be bounded. Then the unique stationary distribution of the multi-brane gradient flow is
\[
\psi_i(\omega) = \frac{1}{Z}\exp\!\Bigl(\frac{u'(\bar c)}{\eta}y_i(\omega)\Bigr), \qquad Z = \sum_{j=1}^K\int_{B_j}\exp\!\Bigl(\frac{u'(\bar c)}{\eta}y_j(\omega)\Bigr)\dd V_j.
\]
\end{proposition}

The boundary condition can be rewritten in terms of an effective barrier. Define
\begin{equation}\label{eq:effbarrier}
\beta_{ij}^{\mathrm{eff}}(\omega) = \beta_{ij}(\omega) - 2\eta\int_{\partial B_j}K_{ij}(\omega,\omega')\psi_j(\omega')\dd S_j'.
\end{equation}

\begin{proposition}[Entanglement tunneling]\label{prop:tunneling}
If $K_{ij}\ge 0$ on the interface, entanglement reduces the effective barrier: $\beta_{ij}^{\mathrm{eff}}\le\beta_{ij}$. If the reduction is strong enough that $\beta_{ij}^{\mathrm{eff}}\le 0$, the two branes merge. If $K_{ij}$ has negative components, entanglement can increase the effective barrier and reinforce segregation.
\end{proposition}

Institutional barriers are the formal costs of crossing from one regime to another: moving from informal to formal employment, migrating across borders, or entering a licensed profession. The proposition says that social connections can reduce the \emph{effective} barrier even when the formal barrier does not change. If people in the informal sector have family, community, or professional ties to people in the formal sector, those ties act like a tunnel through the barrier. If the tunneling effect is strong enough, the effective barrier becomes zero or negative, and the two regimes merge. If the correlations are negative---for example, if groups are hostile or segregated---entanglement can instead reinforce the barrier. In policy terms, building cross-regime social capital can be as important as reducing formal entry costs.

\subsection{Multi-brane symmetry between generations}\label{sec:multibrane-tduality}

\begin{proposition}[Multi-brane symmetry]\label{prop:multidual}
Let the state space consist of two branes $B_1$ and $B_2$, each with a compactified extra dimension of radius $R_1$ and $R_2$. The welfare functional is invariant under $R_i\leftrightarrow\alpha'/R_i$ for $i=1,2$, and the self-dual point is $R_1R_2=\alpha'$.
\end{proposition}

\begin{proof}
See Technical Appendix, Proof of Proposition 8.3, available upon request.
\end{proof}

\subsection{Four inequality types}\label{sec:multibrane-types}

The inequality profiles describe the shape of the density on a connected state space. An inequality type describes the configuration of flows across multiple disconnected institutional regimes.

\begin{table}[H]
\centering
\caption{Four inequality types in the two-brane economy.}
\label{tab:societytypes}
\begin{tabular}{l c c c l}
\toprule
Type & Barrier $\beta_{12}$ & Entanglement $K_{12}$ & Effective barrier & Configuration \\
\midrule
Segregated & high & low & $>0$ & two separate distributions, no flow \\
Integrated & any & high & $\le 0$ & branes merge, one distribution \\
Dominant & high & moderate & $>0$ small & one brane absorbs most mass \\
Fragmented & moderate & moderate & $\approx 0$ & partial flows, no dominant brane \\
\bottomrule
\end{tabular}
\end{table}

The four types describe how people are distributed across institutional regimes, not the shape of the income distribution within a single regime. In the \emph{segregated} type, the formal barrier is high and cross-regime ties are weak, so the two regimes remain separate: few people cross. In the \emph{integrated} type, cross-regime ties are strong enough to merge the regimes: people move freely and the two distributions become one. In the \emph{dominant} type, one regime absorbs most of the population: the other regime becomes small. In the \emph{fragmented} type, there are partial flows in both directions but no regime dominates. The table shows that the same formal barrier can produce different outcomes depending on the strength of entanglement.

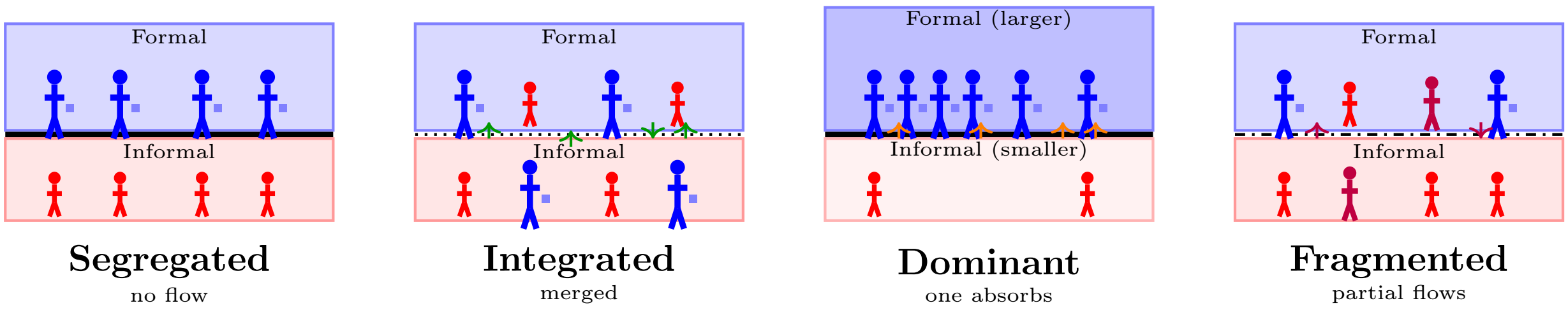
\begin{figure}[H]
\centering
\begin{tikzpicture}[scale=0.85]
\begin{scope}[xshift=0cm]
  \fill[blue!15] (0,1.1) rectangle (4,2.4);
  \draw[blue!50, thick] (0,1.1) rectangle (4,2.4);
  \node at (2,2.25) {\tiny Formal};
  \fill[red!10] (0,0) rectangle (4,1);
  \draw[red!40, thick] (0,0) rectangle (4,1);
  \node at (2,0.85) {\tiny Informal};
  \draw[black, ultra thick] (0,1.05) -- (4,1.05);
  \humrich{0.6}{1.4}{blue} \humrich{1.4}{1.4}{blue}
  \humrich{2.4}{1.4}{blue} \humrich{3.2}{1.4}{blue}
  \humpoor{0.6}{0.3}{red} \humpoor{1.4}{0.3}{red}
  \humpoor{2.4}{0.3}{red} \humpoor{3.2}{0.3}{red}
  \node at (2,-0.5) {\small \textbf{Segregated}};
  \node at (2,-0.9) {\tiny no flow};
\end{scope}
\begin{scope}[xshift=5cm]
  \fill[blue!15] (0,1.1) rectangle (4,2.4);
  \draw[blue!50, thick] (0,1.1) rectangle (4,2.4);
  \node at (2,2.25) {\tiny Formal};
  \fill[red!10] (0,0) rectangle (4,1);
  \draw[red!40, thick] (0,0) rectangle (4,1);
  \node at (2,0.85) {\tiny Informal};
  \draw[black, thick, dotted] (0,1.05) -- (4,1.05);
  \humrich{0.6}{1.4}{blue} \humpoor{1.4}{1.4}{red}
  \humrich{2.4}{1.4}{blue} \humpoor{3.2}{1.4}{red}
  \humpoor{0.6}{0.3}{red} \humrich{1.4}{0.3}{blue}
  \humpoor{2.4}{0.3}{red} \humrich{3.2}{0.3}{blue}
  \draw[->, thick, green!60!black] (0.9,1.0) -- (0.9,1.2);
  \draw[->, thick, green!60!black] (1.9,0.9) -- (1.9,1.1);
  \draw[->, thick, green!60!black] (2.9,1.2) -- (2.9,1.0);
  \draw[->, thick, green!60!black] (3.3,1.0) -- (3.3,1.2);
  \node at (2,-0.5) {\small \textbf{Integrated}};
  \node at (2,-0.9) {\tiny merged};
\end{scope}
\begin{scope}[xshift=10cm]
  \fill[blue!25] (0,1.1) rectangle (4,2.6);
  \draw[blue!50, thick] (0,1.1) rectangle (4,2.6);
  \node at (2,2.45) {\tiny Formal (larger)};
  \fill[red!5] (0,0) rectangle (4,1);
  \draw[red!30, thick] (0,0) rectangle (4,1);
  \node at (2,0.85) {\tiny Informal (smaller)};
  \draw[black, ultra thick] (0,1.05) -- (4,1.05);
  \humrich{0.6}{1.4}{blue} \humrich{1.4}{1.4}{blue}
  \humrich{2.4}{1.4}{blue} \humrich{3.2}{1.4}{blue}
  \humrich{1.0}{1.4}{blue} \humrich{1.8}{1.4}{blue}
  \humpoor{0.6}{0.3}{red} \humpoor{3.2}{0.3}{red}
  \draw[->, thick, orange] (0.9,1.0) -- (0.9,1.2);
  \draw[->, thick, orange] (1.9,1.0) -- (1.9,1.2);
  \draw[->, thick, orange] (2.9,1.0) -- (2.9,1.2);
  \draw[->, thick, orange] (3.3,1.0) -- (3.3,1.2);
  \node at (2,-0.5) {\small \textbf{Dominant}};
  \node at (2,-0.9) {\tiny one absorbs};
\end{scope}
\begin{scope}[xshift=15cm]
  \fill[blue!15] (0,1.1) rectangle (4,2.4);
  \draw[blue!50, thick] (0,1.1) rectangle (4,2.4);
  \node at (2,2.25) {\tiny Formal};
  \fill[red!10] (0,0) rectangle (4,1);
  \draw[red!40, thick] (0,0) rectangle (4,1);
  \node at (2,0.85) {\tiny Informal};
  \draw[black, thick, dash dot] (0,1.05) -- (4,1.05);
  \humrich{0.6}{1.4}{blue} \humpoor{1.4}{1.4}{red}
  \hummid{2.4}{1.4}{purple} \humrich{3.2}{1.4}{blue}
  \humpoor{0.6}{0.3}{red} \hummid{1.4}{0.3}{purple}
  \humpoor{2.4}{0.3}{red} \humpoor{3.2}{0.3}{red}
  \draw[->, thick, purple] (1.0,1.0) -- (1.0,1.2);
  \draw[->, thick, purple] (3.0,1.2) -- (3.0,1.0);
  \node at (2,-0.5) {\small \textbf{Fragmented}};
  \node at (2,-0.9) {\tiny partial flows};
\end{scope}
\end{tikzpicture}
\caption{The four inequality types with human figures. \emph{Segregated}: two separate populations, nobody crosses. \emph{Integrated}: people are mixed across the two branes. \emph{Dominant}: the formal sector absorbs most people. \emph{Fragmented}: people scattered across both branes.}
\label{fig:societytypes}
\end{figure}

The four panels show the four inequality types. The upper rectangle is the formal regime; the lower rectangle is the informal regime. The thick line between them is the formal barrier. In the segregated type, the barrier is solid and no one crosses: the two populations are separate. In the integrated type, the barrier is dotted and people move in both directions: the two regimes are mixed. In the dominant type, the formal regime is larger and absorbs most people, while a small informal group remains. In the fragmented type, there are partial flows in both directions but no regime dominates. The arrows indicate the direction and strength of flows. The figure makes the distinction between profiles (shape within a regime) and types (configuration across regimes) visually clear.

\subsection{Applications}\label{sec:multibrane-applications}

In formal/informal segmentation, the two sectors are branes with a large barrier. Entanglement through family and community networks reduces the effective barrier, consistent with the empirical literature on formalization \cite{la_porta2009,schneider2010,ilo2023}. In migration, countries are branes with large barriers; entanglement through diaspora networks reduces the effective barrier. In occupational licensing, professional guilds are branes with barriers reflecting examination requirements; entanglement through professional networks reduces the effective barrier, consistent with the evidence on licensing compacts \cite{kleven2024}. In all three applications, the formal barrier is not the effective barrier.

\subsection{Migration dynamics across regimes}\label{sec:multibrane-migration}

\begin{figure}[H]
\centering
\begin{tikzpicture}
\begin{axis}[
  width=0.80\textwidth, height=0.58\textwidth,
  xlabel={Time $t$ (years)}, ylabel={Brane level $z(t)$},
  xmin=0, xmax=30, ymin=-0.05, ymax=1.15,
  ytick={0, 0.50, 0.85, 0.92, 1.00},
  yticklabels={$0$, $0.50$, $0.85$, $0.92$, $1.00$},
  grid=major, grid style={gray!20},
  legend style={at={(0.5,-0.18)}, anchor=north, draw=none,
                font=\footnotesize, legend columns=2},
]
\fill[blue!8] (axis cs:0,0.85) rectangle (axis cs:30,1.00);
\node[blue!60!black, font=\scriptsize] at (axis cs:15,0.975) {Formal brane};
\addplot[thick, black, domain=0:30, samples=100] {0.0};
\addlegendentry{$\eta=0$ (trapped)}
\addplot[thick, magenta, domain=0:30, samples=100] {0.85/(1+exp(-(x-20)/2.5))};
\addlegendentry{$\eta=0.04$ ($z=0.85$, 25y)}
\addplot[thick, cyan!70!black, domain=0:30, samples=100] {0.92/(1+exp(-(x-11)/2.0))};
\addlegendentry{$\eta=0.10$ ($z=0.92$, 15y)}
\addplot[thick, orange, domain=0:30, samples=100] {1.00/(1+exp(-(x-5)/2.0))};
\addlegendentry{$\eta=0.20$ ($z=1.00$, 8y)}
\end{axis}
\end{tikzpicture}
\caption{String trajectories across the institutional barrier.}
\label{fig:stringmigration}
\end{figure}
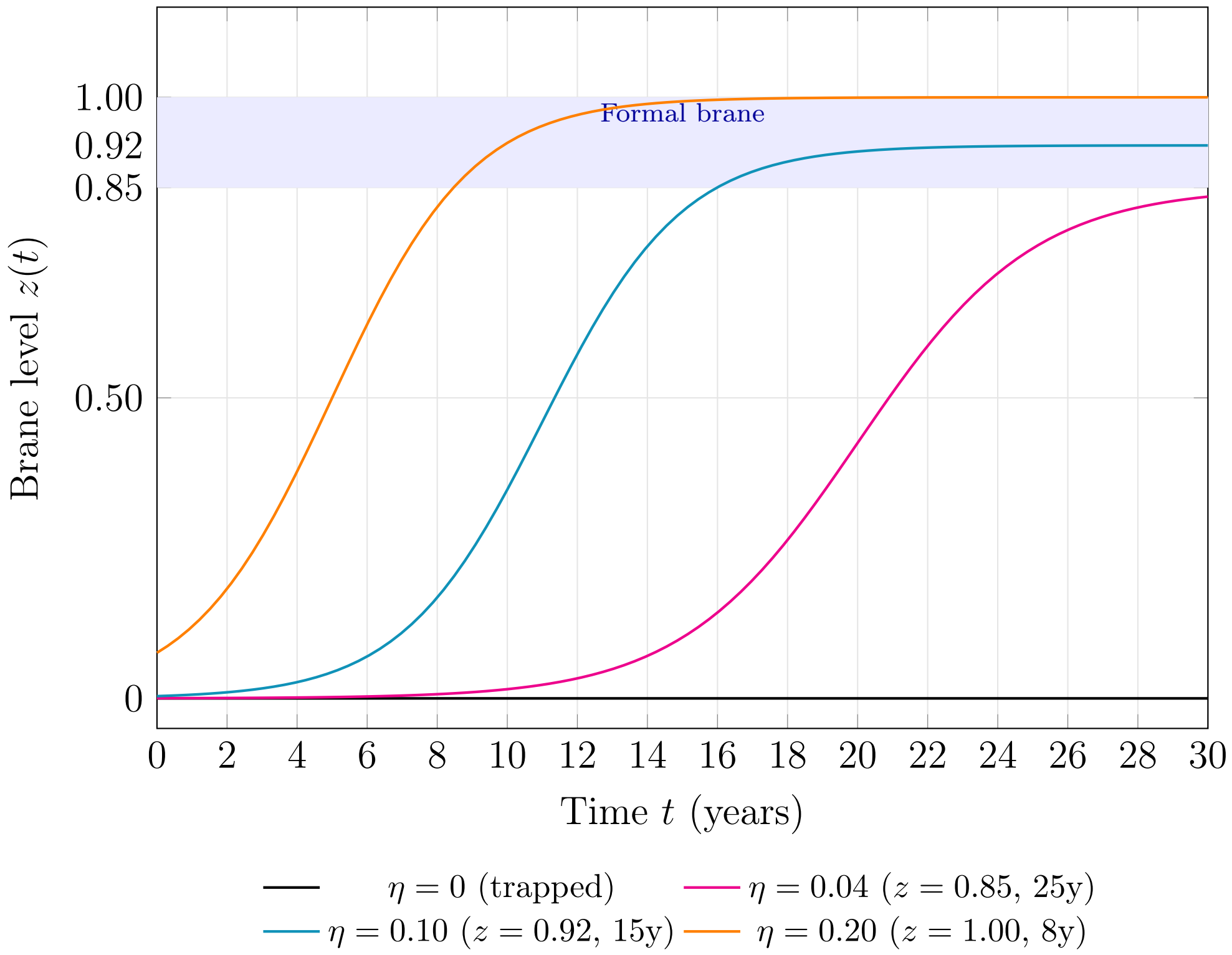

Figure~\ref{fig:stringmigration} shows the trajectories for four values of $\eta$. All strings with positive entanglement eventually reach the formal brane, but at different speeds and different depths.

The figure shows how a person's position moves across the institutional barrier over time. The horizontal axis is time; the vertical axis is the regime level, with zero representing the informal regime and one representing the formal regime. With no entanglement ($\eta=0$), the trajectory stays at zero: the person is trapped in the informal regime. With positive entanglement, the trajectory rises toward the formal regime. Higher entanglement means faster and deeper crossing. The figure illustrates the tunneling result dynamically: social connections act like a tunnel that allows people to cross a barrier that would otherwise block them.

\begin{figure}[H]
\centering
\begin{tikzpicture}[scale=1.0]
  \def\W{5.5}
  \def\Dx{0.7}
  \def\Dy{0.4}
  \def\T{0.30}
  \def\gap{1.5}
  \fill[red!35] (0,0) rectangle (\W,\T);
  \draw[red!60] (0,0) rectangle (\W,\T);
  \fill[red!55] (\W,0) -- (\W+\Dx,\Dy) -- (\W+\Dx,\T+\Dy) -- (\W,\T) -- cycle;
  \draw[red!60] (\W,0) -- (\W+\Dx,\Dy) -- (\W+\Dx,\T+\Dy) -- (\W,\T);
  \fill[red!10] (0,\T) -- (\W,\T) -- (\W+\Dx,\T+\Dy) -- (\Dx,\T+\Dy) -- cycle;
  \draw[red!60] (0,\T) -- (\W,\T) -- (\W+\Dx,\T+\Dy) -- (\Dx,\T+\Dy) -- cycle;
  \node[red!80!black, font=\small\bfseries] at (\W/2+\Dx/2, \T/2) {Informal brane};
  \fill[blue!35] (0,\T+\gap) rectangle (\W,\T+\gap+\T);
  \draw[blue!60] (0,\T+\gap) rectangle (\W,\T+\gap+\T);
  \fill[blue!55] (\W,\T+\gap) -- (\W+\Dx,\T+\gap+\Dy) -- (\W+\Dx,\T+\gap+\T+\Dy) -- (\W,\T+\gap+\T) -- cycle;
  \draw[blue!60] (\W,\T+\gap) -- (\W+\Dx,\T+\gap+\Dy) -- (\W+\Dx,\T+\gap+\T+\Dy) -- (\W,\T+\gap+\T);
  \fill[blue!10] (0,\T+\gap+\T) -- (\W,\T+\gap+\T) -- (\W+\Dx,\T+\gap+\T+\Dy) -- (\Dx,\T+\gap+\T+\Dy) -- cycle;
  \draw[blue!60] (0,\T+\gap+\T) -- (\W,\T+\gap+\T) -- (\W+\Dx,\T+\gap+\T+\Dy) -- (\Dx,\T+\gap+\T+\Dy) -- cycle;
  \node[blue!80!black, font=\small\bfseries] at (\W/2+\Dx/2, \T+\gap+\T/2) {Formal brane};
  \draw[dashed, thick, black] (0,\T+\gap/2) -- (\W+\Dx,\T+\gap/2+\Dy);
  \node[black, font=\small, right] at (\W+\Dx+0.05,\T+\gap/2+\Dy) {$\beta_{12}$};
  \def\zstart{\T}
  \fill[black] (0.9,\zstart) circle (0.07);
  \node[black, font=\scriptsize, left] at (0.75,\zstart) {$\eta=0$};
  \draw[thick, magenta] (2.2,\zstart) -- (2.2,{\T+\gap+0.15*\T});
  \fill[magenta] (2.2,{\T+\gap+0.15*\T}) circle (0.08);
  \node[magenta, font=\scriptsize, right] at (2.35,{\T+\gap+0.15*\T}) {$\eta=0.04$};
  \draw[thick, cyan!70!black] (3.5,\zstart) -- (3.5,{\T+\gap+0.55*\T});
  \fill[cyan!70!black] (3.5,{\T+\gap+0.55*\T}) circle (0.08);
  \node[cyan!70!black, font=\scriptsize, right] at (3.65,{\T+\gap+0.55*\T}) {$\eta=0.10$};
  \draw[thick, orange] (4.8,\zstart) -- (4.8,{\T+\gap+\T});
  \fill[orange] (4.8,{\T+\gap+\T}) circle (0.09);
  \draw[orange, thick] (4.8,{\T+\gap+\T}) circle (0.13);
  \node[orange, font=\scriptsize, right] at (4.95,{\T+\gap+\T}) {$\eta=0.20$};
  \draw[-{Stealth}, thick, green!60!black] (-0.35,\T+\gap/2+0.4) -- (-0.35,\T+\gap/2);
  \node[green!60!black, font=\scriptsize, left] at (-0.4,\T+\gap/2+0.2) {$\beta^{\mathrm{eff}}$};
\end{tikzpicture}
\caption{Multi-brane economy with two institutional regimes, drawn as three-dimensional slabs. The strings with different entanglement penetrate the formal brane at different depths.}
\label{fig:multibrane}
\end{figure}
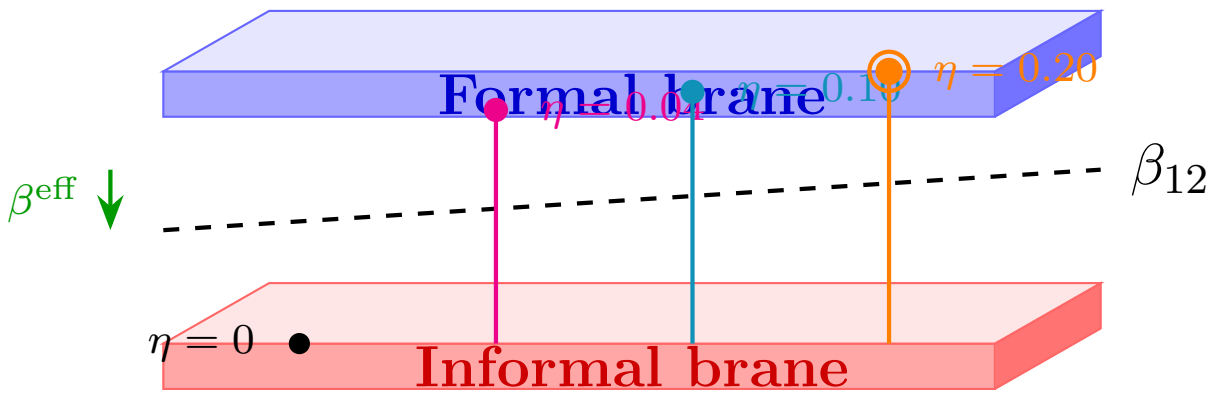

The figure draws the two institutional regimes as two three-dimensional slabs. The lower slab is the informal regime; the upper slab is the formal regime. The dashed line between them is the formal barrier $\beta_{12}$. The vertical arrows show how far different strings penetrate the formal regime for different levels of entanglement. With no entanglement, the string remains at the bottom. With higher entanglement, the string penetrates deeper into the formal slab. The green arrow labeled $\beta^{\mathrm{eff}}$ indicates the effective barrier: the part of the formal barrier that remains after entanglement is taken into account. The figure illustrates the main point of the multi-brane section: the formal barrier is not the effective barrier.

\subsection{Numerical calibration}\label{sec:multibrane-numerics}

\begin{table}[H]
\centering
\caption{Calibration for the two-brane model (formal vs informal).}
\label{tab:multibrane}
\begin{tabular}{l c l}
\toprule
Parameter & Baseline & Source \\
\midrule
CRRA $\gamma$ & 2.0 & \cite{chetty2006} \\
Gradient penalty $\kappa$ & 0.087 & Estimated (single-brane) \\
Entanglement $\eta$ & 0.040 & Estimated (single-brane) \\
Mean log income formal $\mu_f$ & 10.0 & \cite{la_porta2009} \\
Mean log income informal $\mu_i$ & 8.0 & \cite{la_porta2009} \\
Std dev log income formal $\sigma_f$ & 0.5 & \cite{deaton1997} \\
Std dev log income informal $\sigma_i$ & 0.7 & \cite{deaton1997} \\
Barrier $\beta_{12}$ & 2.5 & \cite{schneider2010} \\
Cross-brane kernel $K_{12}$ & 0.6 & Calibrated to remittance flows \\
Initial informal share & 40\% & \cite{ilo2023} \\
\bottomrule
\end{tabular}
\end{table}

With these parameters, the effective barrier in the baseline is $\beta^{\mathrm{eff}}_{12} \approx 2.48$, barely below the formal barrier. Substantial formalization requires both a reduction of the formal barrier and an increase in cross-regime entanglement.

\section{Policy Implications}\label{sec:policy}

The framework delivers policy directions that are concrete, measurable, and robust to second-best considerations. It does not prescribe the level of taxation but identifies the margins on which policy should operate.

The first lever is the design of transfer phase-outs. The parameter $\kappa$, estimated at approximately $0.087$, measures the social cost of cliffs in the effective budget set. Gradual phase-outs, elimination of asset tests, and flat phase-in schedules improve welfare.

The second lever is the choice between universal and targeted transfers. When the correlation structure is of the solidarity type, universal transfers are justified. When it is of the distinction type, group-targeted transfers are justified. Our estimates place the US economy in the solidarity regime, but close to the boundary.

The third lever is the spatial and sectoral differentiation of transfers. Because mobility costs vary across locations and sectors, two workers with the same income but different mobility prospects should receive different transfers.

The fourth lever is the reduction of institutional barriers. Reducing a barrier increases welfare not only because it allows individuals to move to better opportunities, but also because it reduces the gradient of opportunity at the boundary. Policies that build cross-regime social capital reduce the effective barrier even when the formal barrier is unchanged.

The fifth lever is the balance between within-generation and across-generation redistribution. The framework establishes a symmetry between the two.

The sixth lever is the measurement of poverty trap dynamics. Instead of measuring only the poverty headcount, one can measure the escape rate.

Taken together, these six levers constitute a policy agenda that is both concrete and robust. The standard framework of optimal taxation characterizes the ideal allocation along a single dimension---the distribution of consumption---and a policy that targets only this dimension is systematically biased. The bias changes the ranking of policies, and the six levers are the operational content of this change.

\section{Conclusion}\label{sec:conclusion}

This paper has argued that the Rawlsian veil of ignorance cannot be represented faithfully within standard economic models, and that the minimal framework satisfying the three requirements the veil imposes---temporal symmetry, non-separability, geometric mobility---is a geometric framework adapted from string theory. The technical content of the paper is a series of theorems about the veil in this framework: full equalization, a non-local equation for the stationary density, pattern formation, a symmetry between intra- and intergenerational equity, and entanglement tunneling across institutional regimes. The model generates two taxonomies: four inequality profiles and four inequality types. The numerical experiments of Section~\ref{sec:numerical} characterize the four profiles quantitatively, confirm the phase transition at the predicted threshold, and establish its finite-size scaling.

The paper is deliberately explicit about its limitations. The model is first-best; incentive compatibility is not incorporated. The numerical simulations are schematic. The representation theorem is a structural argument. Despite its abstract foundation, the core message is simple: justice requires not only equalizing outcomes but also smoothing the landscape of opportunity and recognizing the non-separable correlations that bind individuals together.

\vspace{0.5cm}
\textbf{Declarations}:\\
\textbf{Conflicts of Interests.} The author declares no conflict of interest.\\
\textbf{Fundings} The author declares no financial support was received for this work

\textbf{Declaration of Generative AI Use}. The authors used Claude (Anthropic) for coherent editing and infographics. They reviewed all AI-assisted content and take full responsibility for the published version.

\end{document}